\documentclass[12pt,draftcls,onecolumn]{IEEEtran}

\usepackage{graphicx, array, bm, setspace, amsmath, amsfonts, amssymb, acronym, citesort, units, verbatim, subfigure, algorithm, algorithmic, color, float, latexsym, epsfig, afterpage, pstricks}

\newcommand{\ie}{\emph{i.e.}, }
\newcommand{\eg}{\emph{e.g.}, }

\newcommand{\vect}[1]{\mathrm{\boldsymbol{#1}}}

\newcommand{\thickhline}{\noalign{\hrule height 0.8pt}}

\allowdisplaybreaks

\acrodef{ms}[MS]{mobile station}
\acrodef{bs}[BS]{base station}
\acrodef{rb}[RB]{resource block}
\acrodef{ap}[AP]{access point}
\acrodef{hetnet}[HetNet]{Heterogeneous Network}
\acrodef{ofdma}[OFDMA]{orthogonal frequency division multiple access}
\acrodef{icic}[ICIC]{inter-cell interference coordination}
\acrodef{sinr}[SINR]{signal-to-interference-plus-noise ratio}
\acrodef{cci}[CCI]{Co-channel interference}
\acrodef{cc}[CC]{component carrier}
\acrodef{mcs}[MCS]{modulation and coding scheme}
\acrodef{fbs}[FBS]{femto-base station}
\acrodef{abs}[ABS]{Almost-blank Subframe}
\acrodef{lte}[LTE]{Long-Term Evolution}
\acrodef{ltea}[LTE-A]{\ac{lte}-Advanced}
\acrodef{rsrp}[RSRP]{reference signal received power}
\acrodef{4g}[4G]{4$^{\rm th}$ generation}
\acrodef{cdf}[CDF]{cumulative distribution function}
\acrodef{ffr}[FFR]{fractional frequency reuse}
\acrodef{dfr}[DFR]{dynamic frequency reuse}
\acrodef{cdf}[CDF]{cumulative distribution function}
\acrodef{pdf}[PDF]{probability distribution function}
\acrodef{pmf}[PMF]{probability mass function}
\acrodef{la}[LA]{link adaptation}
\acrodef{pfs}[PFS]{proportional fair scheduler}
\acrodef{popc}[POPC]{Pareto optimal power control}
\acrodef{minlp}[MINLP]{mixed-integer non-linear programming}
\acrodef{son}[SON]{Self-Organising Network}
\acrodef{rntp}[RNTP]{relative narrowband transmit power}

\newcommand{\Hhuge}{\fontsize{23}{26}\selectfont}

\begin{document}


\title{\Hhuge \vspace*{-30pt} Distributed and Autonomous Resource and Power Allocation for Wireless Networks}

\author{
Harald Burchardt\authorrefmark{1}, Sinan Sinanovi\'c\authorrefmark{1}, Zubin Bharucha\authorrefmark{2} and Harald Haas\authorrefmark{1}\\
 \bigskip
\begin{minipage}{8.1cm}
 \begin{center}
{\authorrefmark{1}\normalsize  Institute for Digital Communications}\\
{\normalsize  School of Engineering and Electronics}\\
{\normalsize  The University of Edinburgh}\\
{\normalsize EH9 3JL, Edinburgh, UK}\\
{\normalsize   \{h.burchardt, s.sinanovic, h.haas@ed.ac.uk\} \vspace*{-4mm}}\vspace*{0.0cm}
 \end{center}
 \end{minipage}
       \hspace*{-0.0cm}
 \begin{minipage}{8.1cm}
 \begin{center}
{\authorrefmark{2}\normalsize DOCOMO Euro-Labs}\\
{\normalsize Landsbergerstr. 312}\\ 
{\normalsize 80687 Munich, Germany}\\
{\normalsize bharucha@docomolab-euro.com \vspace*{-3mm}}
 \end{center}
 \end{minipage}
}
\maketitle

\vspace*{-35pt}
\begin{abstract}\vspace*{-10pt}

In this paper, a distributed and autonomous technique for resource and power allocation in \ac{ofdma} femto-cellular networks is presented. Here, \acp{rb} and their corresponding transmit powers are assigned to the user(s) in each cell individually without explicit coordination between \acp{fbs}. The ``allocatability'' of each resource is determined utilising only locally available information of the following quantities:
\begin{itemize}
 \item the required rate of the user;
 \item the quality (\ie strength) of the desired signal;
 \item the level of interference incident on each \ac{rb}; and
 \item the frequency-selective fading on each \ac{rb}.
\end{itemize}
Using a fuzzy logic system, the time-averaged values of each of these inputs are combined to determine which \acp{rb} are most suitable to be allocated in a particular cell, \ie which resources can be assigned such that the user requested rate(s) in that cell are satisfied. Furthermore, \ac{la} is included, enabling users to adjust to varying channel conditions. A comprehensive study 
in a femto-cell environment is performed, yielding system performance improvements in terms of throughput, energy efficiency and coverage over state-of-the-art \ac{icic} techniques.

\end{abstract}

\vspace*{-15pt}\begin{keywords}\vspace*{-10pt}
autonomous resource allocation, distributed \ac{icic}, fuzzy logic, \ac{ofdma}, femto-cellular networks.
\end{keywords}
\vspace*{-15pt}
\acresetall
\section{Introduction}
\label{sec:intro}



Future wireless networks are moving towards heterogeneous architectures, where in each cell a user may have over four different types of \acp{ap} (\eg macro-, pico-, femto-cells, relays and/or remote radio heads)~\cite{lgrkqz1101}. Intuitively, this has many positive effects for a \ac{ms}, which can now choose from several \acp{bs} to find the most suitable. However, pico- and femto-cellular overlays also imbue many difficulties, \eg cell-organisation/optimisation, resource assignment to users, and especially interference coordination between \acp{ap} within the same and neighbouring cells. Standard \ac{icic} techniques based on network architectures~\cite{bpgcwv0901,adfjlp0901} only go so far in dealing with these challenges, and hence a new approach is necessary.

\vspace*{-5pt}
\subsection{Challenges in \acp{hetnet}}

Through the various types, locations and dense deployment of \acp{ap}, and the different transmissions powers/ranges associated with them, numerous technical challenges are posed by femto/pico-cell overlays~\cite{lgrkqz1101,mbsbkj1001,clcc1101}. These mainly fall into the following areas:
\begin{itemize}
 \item \textbf{Network self-organisation} - Self-configuration and -optimisation are required of all cells. In cellular networks, such organisation can be performed via optimisation techniques~\cite{hkg0901}, however these tasks become increasingly difficult given the additional \acp{ap} and network parameters to be considered, motivating a \emph{distributed} configuration approach~\cite{gkgo0701}.
 \item \textbf{Backhauling} - Connecting the different \acp{bs} to the core-network necessitates extra infrastructure~\cite{lgrkqz1101}. In the femto-cell case, the long delay of connection via wired backhaul prevents macro-femto \ac{icic}~\cite{clcc1101}, and hence necessitates \emph{autonomous} interference management.
 \item \textbf{Interference} - Cross-tier interference created to/from the overlaid cells (\eg pico-/femto-cells) must be mitigated to maintain performance, especially if access to these cells is restricted. High intra-femto-tier interference due to dense deployment is also of concern. The handling of this interference is paramount to the performance of such future networks, of which the main sources in densely deployed femto-cell scenarios~\cite{lgrkqz1101} can be given as
\begin{itemize}
 \item \emph{Unplanned deployment}-  Low-power nodes such as femto-cells are deployed by end-users at ``random'' locations, and can be active or inactive at any time, further randomising their interference. Continuous sensing and monitoring is required by cells to dynamically/adaptively mitigate interference from the other tiers~\cite{lmssn0901}.
 \item \emph{Closed-subscriber access} - Restricted access control of pico- and femto-cells may lead to strong interference scenarios in downlink and uplink if users cannot handover.
 \item \emph{Node transmission power differences} - The lower power of nodes such as pico- and femto-cells can cause associations downlink/uplink interference problems.
\end{itemize}
\end{itemize}

In general, these issues motivate the need for \textbf{\emph{decentralised, autonomous}} interference coordination schemes that operate independently on each cell, utilising only local information, yet achieving efficient/near-optimal solutions for the network. 
By allowing \acp{bs} (all types) and \acp{ms} to individually optimise their resource allocations and transmission powers, a global optimum may be found without centralised algorithms governing the system. This would substantially reduce not only the amount of signalling but also the operation complexity of the network.

\subsection{Randomly and Densely Deployed Femto-cells}

Here, we address the relatively unexplored topic of \ac{icic} for randomly deployed femto-cells. Due to the relative modernity of the femto-cell concept, and the innate random deployment of femto-cells within a macro-cell, most interference coordination techniques are utilised for interference reduction to the macro-cell, rather than interference protection between femto-cells. 



The state-of-the-art interference coordination for \ac{lte} \acp{hetnet} is the \ac{abs}: a time-domain \ac{icic} technique where an aggressor \ac{bs} creates ``protected'' subframes for a victim \ac{bs} by reducing its transmission activity on these~\cite{tech:llkymc1101}; the occurrences of the \acp{abs} are known \textit{a priori} at the coordinating \acp{bs}. Thus, throughput improvements are induced via the provided interference protection~\cite{pwwsjl1201}. However, the omitted transmission frames may have adverse affects on the data rates at the agressor \ac{bs}. Furthermore, without guaranteed backhaul connections, \acp{fbs} may not be able coordinate the \ac{abs} slots. In this paper, we provide resource and power allocation for femto-femto interference environments which requires no signalling between \acp{fbs}, and enhances the overall throughput, energy efficiency and fairness of the femto-network.



On another note, recent research has seen the emergence of autonomous coordination techniques for \acp{son}~\cite{sv0901,caha1101}, where transmit powers on subbands is adjusted independently in each cell via local and network utility optimisation. These utilities are based on the average rate in the cell, however do not consider user-specific resource allocation for additional interference coordination. Furthermore, the proposed strategies do not consider heterogeneous architectures that will inevitably describe future networks. Finally, the suggested algorithms assume still some signalling between neighbouring \acp{bs}, hence cannot be considered fully autonomous, and may also limit their applicability specifically for femto-cell networks.

Finally, the application of fuzzy logic in collaboration with reinforcement learning techniques is comprehensively studied in~\cite{j9801}, in order to tune the outputs of fuzzy inference systems. The application to wireless network coordination is investigated in~\cite{da1001,vttd1101,rkc1001}, where fuzzy logic reduces the complexity of the learning algorithms by providing coarse evaluations of the network state. On a cell-individual basis, by again adapting subband transmission powers ~\cite{da1001}, modifying the downlink \ac{rntp} thresholds~\cite{vttd1101}, or adjusting the antenna downtilt~\cite{rkc1001} the interference on specific resources can be controlled or removed completely, respectively. On the other hand, QoS requirements of individual users are neglected, a perspective that we attempt to address here. In addition, we employ a holistic approach by considering many key parameters to perform resource allocation (\ie frequency reuse) and power control in all cells individually.

In this paper, we introduce a novel, low-complexity, distributed and autonomous \ac{icic} technique, that performs independent resource and power allocation in each cell, eliminating explicit signalling between \acp{fbs}. The rest of the paper is structured as follows: Section~\ref{sec:sysmodel} describes the system deployment scenario and channel environment, Section~\ref{sec:fuzzylogic} explains the fuzzy logic \ac{icic} protocol and its performance in femto-cellular networks is analysed in Section~\ref{sec:optim}. In Section~\ref{sec:simulator} the simulation is described, and Section~\ref{sec:results} portrays and discusses the simulation results. Finally, some concluding remarks are offered in Section~\ref{sec:concl}.

\section{System and Channel Model}
\label{sec:sysmodel}

An \ac{ofdma} network is considered, where the system bandwidth $B$ is divided into $M$ \acp{rb}. A \ac{rb} defines one basic time-frequency unit of bandwidth $B_{\rm RB}{=}\nicefrac{B}{M}$. All \acp{ms} can transmit up to a fixed maximum power $P_{\max}$. Perfect time and frequency synchronisation is assumed.

Universal frequency reuse is considered, such that each femto-cell utilises the entire system bandwidth $B$. The set of \acp{rb} $\mathcal{M}$, where ${\left\lvert \mathcal{M}\right\rvert{=}M}$, is distributed by each \ac{bs} to its associated \ac{ms}(s). Throughout this paper, $u$ is used to define any \ac{ms}, and $v_u$ the \ac{bs} with which this \ac{ms} is associated. The received signal observed by \ac{ms}$_u$ from \ac{bs}$_{v_u}$ on \ac{rb} $m$ is given by
\begin{equation}
Y^m_u = \underbrace{P_u^m G^m_{u,v_u}}_{S^m_u} + I^m_u + \eta\,,
\end{equation}
where $G^m_{u,v_u}$ signifies the channel gain between the \ac{ms}$_u$ and its serving \ac{bs}$_{v_u}$, observed on \ac{rb} $m$. Furthermore, $P_u^m$ denotes the transmit power of \ac{ms}$_u$ on \ac{rb} $m$, $S^m_u$ the desired received signal, $\eta{=}\eta_0 B_{\rm RB}$ the thermal noise, and $I^m_u$ the co-channel interference received on \ac{rb} $m$ from \acp{ms} in neighbouring cells. The interference $I^m_u$ is defined by
\begin{equation}
I^m_u = \sum_{i \in \mathcal{I}} P_i^m G^m_{u,v_i}\,,
\end{equation}
where $\mathcal{I}$ represents the set of interferers (\ie set of \acp{ms} in neighbouring cells that are also assigned \ac{rb} $m$). Hence, the \ac{sinr} observed at the \ac{ms}$_u$ on \ac{rb} $m$ is calculated by
\begin{equation}
\gamma^m_u = \frac{S^m_u}{I^m_u + \eta} = \frac{P_u^m G^m_{u,v_u}}{\displaystyle\sum_{i \in \mathcal{I}} P_i^m G^m_{u,v_i} + \eta}\,.
\end{equation}
Following this, the user throughput $C_u$ is calculated as the data transmitted on the assigned \acp{rb} that have achieved their \ac{sinr} target $\gamma^*_u$
\begin{equation}
C_u = \tilde{n}^{\rm RB}_u k_{\rm sc} s_{\rm sc} \varepsilon_{\rm s}\,,
\label{eq:C_u}
\end{equation}
where $\tilde{n}^{\rm RB}_u{=}\sum^{n^{\rm RB}_u}_{m{=}1}\vect{1}_{\gamma_u^m\geq\gamma^*_u}$ is the number of \acp{rb} assigned to \ac{ms}$_u$ achieving $\gamma^*$, $n^{\rm RB}_u$ is the total number of \acp{rb} allocated to \ac{ms}$_u$, $\vect{1}_A$ is the indicator function, $k_{\rm sc}$ the number of subcarriers per \ac{rb}, $s_{\rm sc}$ the symbol rate per subcarrier, and $\varepsilon_{\rm s}$ the \ac{mcs} 
given in Table~\ref{tab:amc}\footnote{In Table~\ref{tab:amc}, the modulation and coding orders are taken from \ac{lte}~\cite{book:stb0901}, and the \ac{sinr} ranges from~\cite{tech:e1001}. In general, these values are operator specific, and hence are not standardised.}.
\begin{table}[htb]
\centering
\caption{Modulation and Coding Table}
\begin{tabular}{c|c|ccc}
\thickhline
CQI & min. &  & Code & Efficiency \\
index & SINR [dB] & Modulation & rate & $\varepsilon_{\rm s}$ [bits/sym] \\
\hline
0 & - & None & - & 0 \\
1 & -6 & QPSK & 0.076 & 0.1523 \\
2 & -5 & QPSK & 0.12 & 0.2344 \\
3 & -3 & QPSK & 0.19 & 0.3770 \\
4 & -1 & QPSK & 0.3 & 0.6016 \\
5 & 1 & QPSK & 0.44 & 0.8770 \\
6 & 3 & QPSK & 0.59 & 1.1758 \\
7 & 5 & 16QAM & 0.37 & 1.4766 \\
8 & 8 & 16QAM & 0.48 & 1.9141 \\
9 & 9 & 16QAM & 0.6 & 2.4063 \\
10 & 11 & 64QAM & 0.45 & 2.7305 \\
11 & 12 & 64QAM & 0.55 & 3.3223 \\
12 & 14 & 64QAM & 0.65 & 3.9023 \\
13 & 16 & 64QAM & 0.75 & 4.5234 \\
14 & 18 & 64QAM & 0.85 & 5.1152 \\
15 & 20 & 64QAM & 0.93 & 5.5547 \\
\thickhline
\end{tabular}
\label{tab:amc}
\end{table}
Finally, the system capacity is calculated as the sum throughput of all \acp{ms} in the network
\begin{equation}
C_{\rm sys} = \sum_u C_u\,.
\label{eq:c_sys_def}
\end{equation}

The power efficiency $\beta_u$ measures the data rate per unit of transmit power (or, alternatively, the data sent per unit of energy) of \ac{ms}$_u$. This is defined as follows:
\begin{equation}
\beta_u = \frac{C_u}{P_u} = \frac{ \tilde{n}^{\rm RB}_u k_{\rm sc} s_{\rm sc} \varepsilon_{\rm s}}{ \sum^{n^{\rm RB}_u}_m P^m_u} \quad {\rm \left[\frac{bits/s}{W}\right]\equiv\left[\frac{bits}{J}\right]}\,,
\label{eq:peff}
\end{equation}
where $P_u$ is the transmit power of \ac{ms}$_u$, and $C_u$ the achievable capacity from~\eqref{eq:C_u}. 
The availability $\chi$ is defined as the proportion of \acp{ms} that have acquired their desired rate, \ie 
\begin{equation}
\chi{=}\frac{1}{n_{\rm usr}}\sum_{u=1}^{n_{\rm usr}}\vect{1}_{C_u\geq C^*_u}\,,
\end{equation}
where $n_{\rm usr}$ is a random variable denoting the number of \acp{ms} in the scenario 
and $C_u^*$ is the desired rate of \ac{ms}$_u$. Lastly, Jain's Fairness Index~\cite{tech:jch8401} is used to calculate the throughput fairness of the system in each time slot
\begin{equation}
f(\mathbf{C}) = \frac{\left[\sum_u C_u \right]^2}{\sum_u C_u^2}\,,
\label{eq:fairness}
\end{equation}
where the vector $\mathbf{C}$ denotes the achieved throughputs of all \acp{ms} in the system.

\subsection{Scenario Construction}
\label{sec:scenmodel}

\begin{figure}[htb]
\centering
\includegraphics[width=.6\columnwidth]{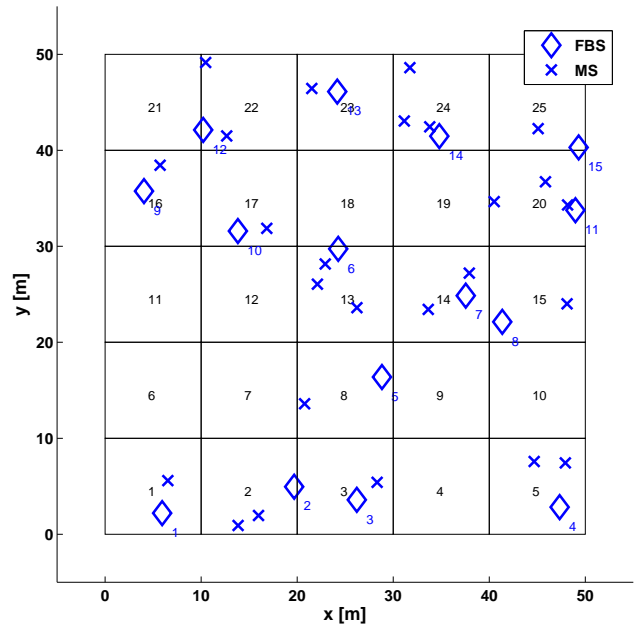}
\caption{Apartment block scenario with ${p_{\rm act}=0.5}$, where each apartment is $10\,{\rm m}{\times}10\,{\rm m}$, with $\tilde{\mu}(u){=}3$ and equal user number probabilities.}
\label{fig:apart_scen}
\end{figure}

A ${5{\times}5}$ apartment grid is considered for the femto-cell scenario, where the probability ${p_{\rm act}}$ describes the likelihood of an active \ac{fbs} in a given apartment. Furthermore, we assume that multiple \acp{ms} may be present in an apartment. 
As it is unlikely all cells will have the same number of \acp{ms}, the user generation is implemented via probability table, 
where depending on the maximum number of users $\tilde{\mu}(u)$ allowed per cell, the number of \acp{ms} $n_c(u){\in}\{1,\dots,\tilde{\mu}(u)\}$ present in cell $c$ is randomly chosen. Table~\ref{tab:prob_table} gives two examples of probability tables, where \emph{a)}~equal probabilities are given to all $n(u)$, or \emph{b)}~the probability reduces with each additional~\ac{ms}. 
\begin{table}[htb]
\centering
\caption{Probability tables for the number of users allocated in a single femto-cell. The left table induces equal probabilities for each possible number of users, in the right table the probability is halved with each additional user.}
\begin{tabular}{c|cccc}
$\tilde{\mu}(u)$ & 1 & 2 & 3 & 4 \\
\hline
$p_{n(u){=1}}$ & 1 & \nicefrac{1}{2} & \nicefrac{1}{3} & \nicefrac{1}{4} \\
$p_{n(u){=2}}$ & 0 & \nicefrac{1}{2} & \nicefrac{1}{3} & \nicefrac{1}{4} \\
$p_{n(u){=3}}$ & 0 & 0 & \nicefrac{1}{3} & \nicefrac{1}{4} \\
$p_{n(u){=4}}$ & 0 & 0 & 0 & \nicefrac{1}{4}
\end{tabular}
\hspace*{15pt}\mbox{\textbf{or}}\hspace*{15pt}
\begin{tabular}{c|cccc}
$\tilde{\mu}(u)$ & 1 & 2 & 3 & 4 \\
\hline
$p_{n(u){=1}}$ & 1 & \nicefrac{2}{3} & \nicefrac{4}{7} & \nicefrac{8}{15} \\
$p_{n(u){=2}}$ & 0 & \nicefrac{1}{3} & \nicefrac{2}{7} & \nicefrac{4}{15} \\
$p_{n(u){=3}}$ & 0 & 0 & \nicefrac{1}{7} & \nicefrac{2}{15} \\
$p_{n(u){=4}}$ & 0 & 0 & 0 & \nicefrac{1}{15}
\end{tabular}
\label{tab:prob_table}
\end{table}
Here, we utilise $\tilde{\mu}(u){=}3$. An example of such a scenario is shown in Fig.~\ref{fig:apart_scen}.
In each active femto-cell, both the \acp{ms} and \ac{fbs} are uniformly distributed in the apartment. Due to the private deployment of femto-cells a closed-access system is assumed~\cite{bhsa1001}, so each \ac{ms} is assigned to the \ac{fbs} in its apartment, even if a foreign cell exhibits superior link conditions.

\subsection{Channel Model}
\label{sec:chanmod}

In general, the channel gain, $G^m_{k,l}$, between a transmitter~$l$ and receiver~$k$, observed on \ac{rb} $m$ and separated by a distance $d$ is determined by the path loss, log-normal shadowing, and channel variations caused by frequency-selective fading:
\begin{equation}
G^m_{k,l} = \left\lvert H^m_{k,l}\right\rvert ^2 10^{\frac{-L_d(d) + X_\sigma}{10}},
\label{eq:channelfunc}
\end{equation}
where $H^m_{k,l}$ describes the channel transfer function between transmitter $l$ and receiver $k$ on \ac{rb} $m$, $L_d(d)$ is the distance-dependent path loss (in dB) and $X_\sigma$ is the log-normal shadowing value (in dB) with standard deviation $\sigma$, as described in~\cite{wd:imt-chanmod}. The channel response generally exhibits time and frequency dispersions, however channel fluctuations within a \ac{rb} are not considered as the \ac{rb} dimensions are significantly smaller than the coherence time and bandwidth of the channel~\cite{wosas0301}. Furthermore, the path loss $L_d(d)$ is identical on all \acp{rb} assigned to the \ac{ms}. Finally, the delay profiles used to generate the frequency-selective fading channel transfer factor $H^m_{k,l}$ are taken from applicable propagation scenarios in~\cite{wd:imt-chanmod},~\cite{std:3gpp-nem}.

The path loss model used to calculate $L_d(d)$ is for indoor links~\cite{std:3gpp-simhenbrfreq}, \ie the link (desired or interfering) between a \ac{fbs} and an indoor \ac{ms}, and calculates the path loss as
\begin{equation}
L_d(d) = \alpha + \beta \log_{10}(d) \quad [\mbox{dB}]\,.
\label{eq:pathlossfunc}
\end{equation}
where $d$ is the distance between transmitter and receiver, and $\alpha,\,\beta$ are the channel parameters.

Log-normal shadowing is added to all links through correlated shadowing maps. These are generated such that the correlation in shadowing values of two points is distance-dependent. Table~\ref{tab:sim_param} shows the shadowing standard deviation $\sigma$ and auto-correlation distances considered~\cite{std:3gpp-simhenbrfreq}.

\section{Distributed and Autonomous Resource Allocation}
\label{sec:fuzzylogic}


Due to the customer-side random deployment of femto-cells, and the resulting lack of fixed connective infrastructure, 
\acp{fbs} must perform resource and power allocation utilising locally available information only. To maximise the performance in its own cell, a \ac{fbs} must attempt to allocate \acp{rb} such that the desired signal on these is maximised, while the interference incident from neighbouring cells is minimal. Furthermore, the \ac{bs} must allocate enough resources such that the rate requirements of the user(s) in the cell are fulfilled. The necessary, and locally available, information is therefore clearly determined:
\begin{itemize}
 \item the required rate of a user determines the number of \acp{rb} that need to be assigned;
 \item the quality (\ie strength) of the desired signal dictates the necessary transmit power;
 \item the level of interference incident on the \acp{rb} strongly influences their allocatability; and
 \item the frequency-selective fading profile also affects the preferable \acp{rb} to be allocated.
\end{itemize}
All of these variables are locally available at the \ac{fbs} in the reverse link, and at the \ac{ms}(s) in the forward link, necessitating no extra information to be exchanged between \acp{bs}. 

\subsection{Fuzzy Logic for Autonomous Interference Coordination}

In general, the resource and power allocation problem for a multi-cellular wireless network belongs to the class of \ac{minlp} problems; obtaining the solutions to these is known to be $\mathcal{NP}$-hard~\cite{as1101,book:b9901}. Therefore, it is clear that a heuristic for local, autonomous resource management is required to solve this problem. A machine learning approach where \acp{fbs} acquire information about their transmission conditions over time would be such a viable solution, however can prove complex without the availability of training data. Therefore, we introduce fuzzy logic as our heuristic, through which ``expert knowledge'' is incorporated in the \ac{rb} allocation decision process. 

\begin{figure}[htb]
\centering
\includegraphics[width=.7\textwidth]{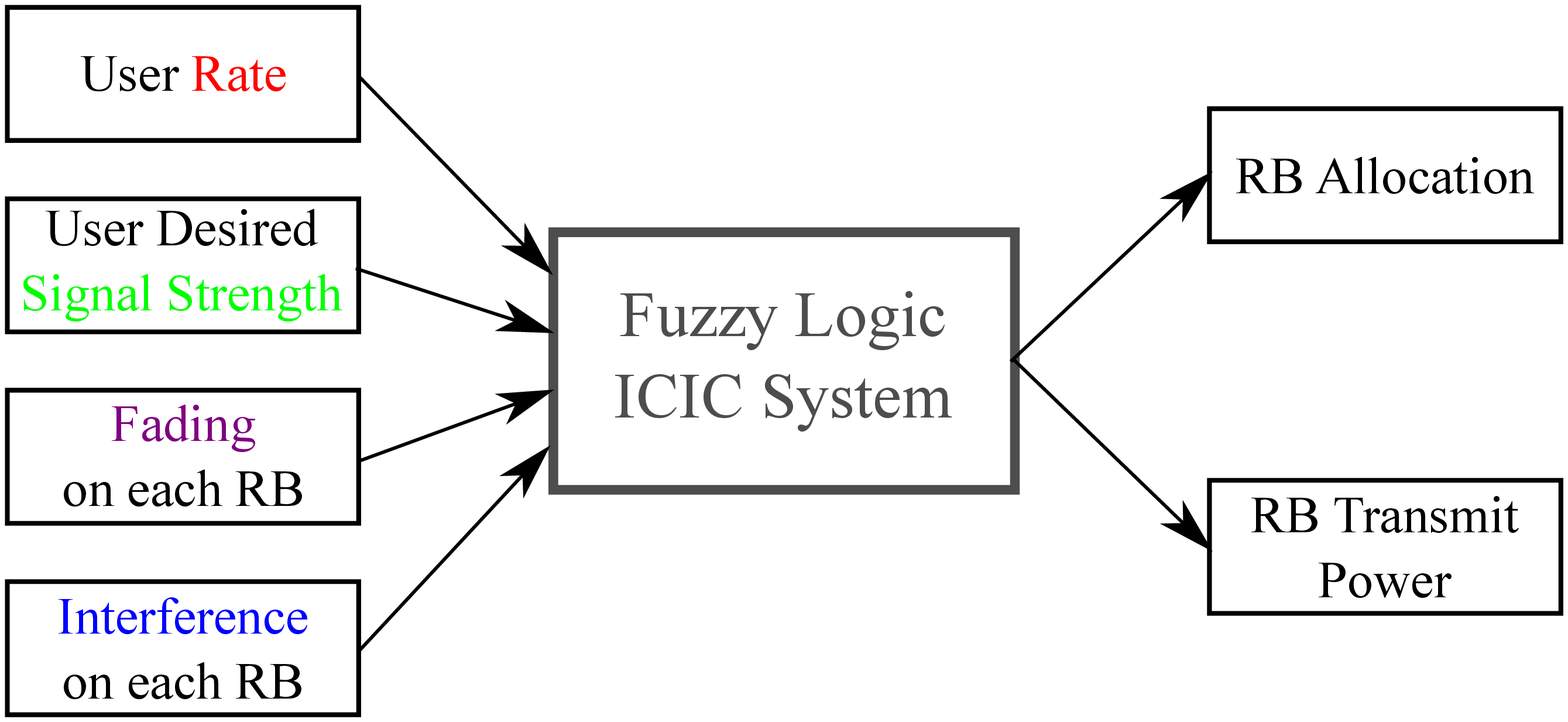}
\caption{Simplified graphical representation of our autonomous resource and power allocation technique.}
\label{fig:simplified_diagram}
\end{figure}

The decision system, in its most simplified form, is represented in Fig.~\ref{fig:simplified_diagram}. In broad terms, the system evaluates which \ac{rb}(s) are most suitable to be allocated to the \ac{ms} in a particular time slot, and determines the transmit power on these \acp{rb} to generate the required \ac{sinr} such that the user's rate can be met. Obviously, an \ac{rb} receiving little or no interference situated in a fading peak is most suitable for allocation to the femto-user, whereas any \ac{rb}(s) receiving high interference, or experiencing deep fades, are much less appropriate.

In fuzzy logic, an input range is divided into multiple ``membership functions'' which give a coarse evaluation of the variable. 
By combining the membership values of the inputs through various rules, the allocatability of each \ac{rb} is determined. The output is also ``fuzzy,'' indicating how suitable (or unsuitable) an \ac{rb} is given the current inputs, avoiding a hard yes/no decision. 

In each time slot, the \ac{fbs} allocates the most applicable \acp{rb} to each \ac{ms}, and data transmission is performed. Based on the received signal levels from the desired user and interfering \acp{ms}, the \ac{bs} updates its information to more accurately represent the long-term interference and fading environments of its cell. This updated information is utilised in the next time slot to again carry out the, hopefully improved, resource and power allocation. The same operation is performed in all femto-cells in the scenario, and the \ac{rb} allocations are continuously individually optimised until the system converges to a stable solution, in which the user(s) in each cell are satisfied.

\subsubsection{Inputs}
\label{sec:fl_inputs}

\begin{figure}[p]
\centering
\includegraphics[width=\textheight,angle=-90]{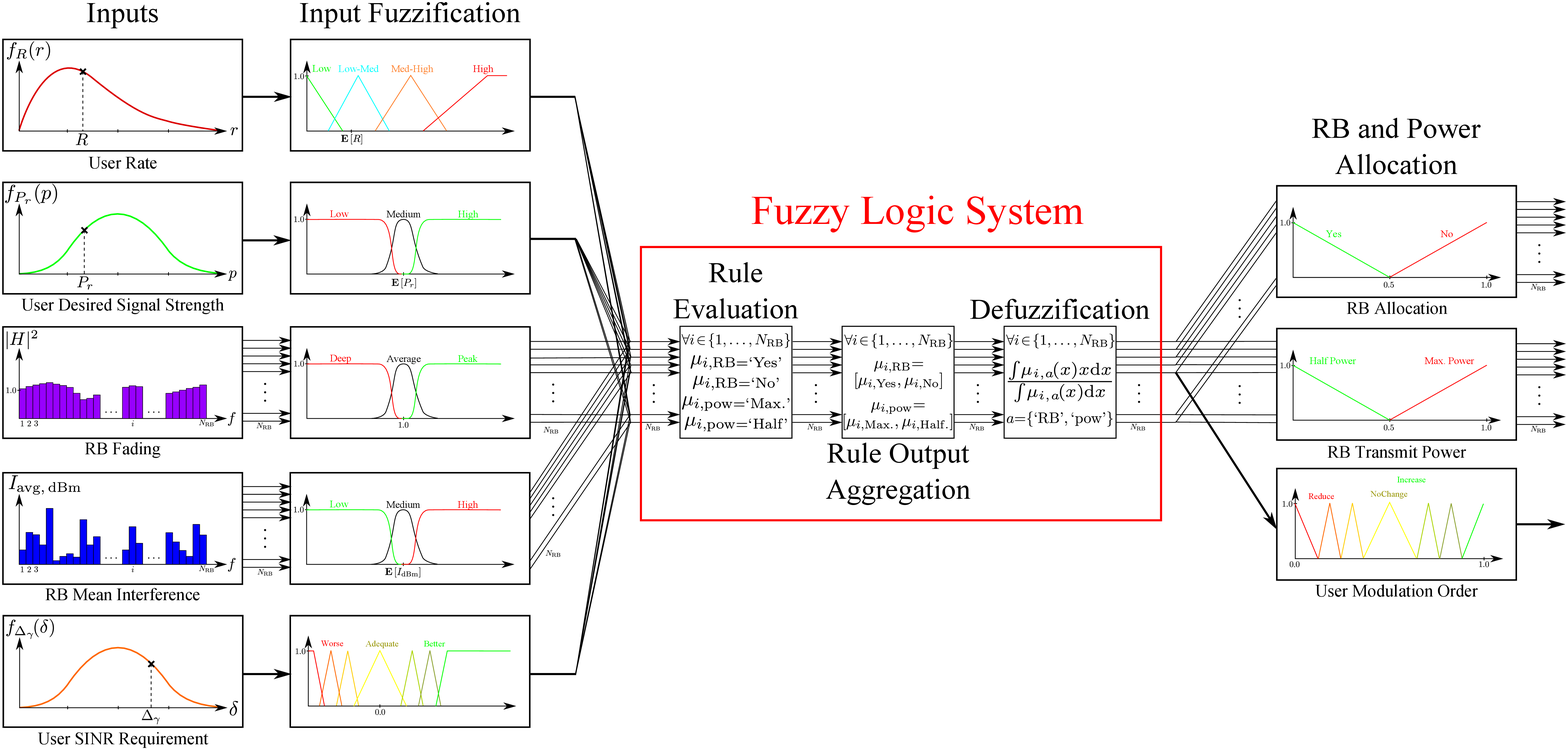}
\caption{Graphical representation of fuzzy logic resource and power allocation system.}
\label{fig:fuzzy_logic_system}
\end{figure}

\noindent The input variables considered in the fuzzy logic system are:
\begin{itemize}
\item The \textbf{required rate} of the \ac{ms} is defined by the service being demanded by the user. 
Here, the values ``Low,'' ``Low-medium,'' ``Medium-high,'' and ``High'' are used to categorise the rate requested by the user. The ranges of these are dependent on the user scenario (\eg in femto-cells, a higher rate can be requested due to the superior channel conditions). This is a per-user requirement, and thus is equivalent for all \acp{rb}.

\item The \textbf{desired signal level} describes the transmission conditions from transmitter to receiver, \ie the stronger the desired signal, the better the channel between the two. 
The signal power domain is divided into ``Low,'' ``Medium,'' and ``High'' values\footnote{The cut off points and slopes of the values are determined from the \acp{cdf} in Fig.~\ref{sfig:sigcdfs}.}, to sort users depending on their useful channels.
Since we consider the fast fading component as a separate input variable, the desired signal level is described per \ac{ms}, and thus is equivalent over all \acp{rb}.

\item The \textbf{level of interference} illustrates the immediate interference environment for each \ac{ms} on each \ac{rb}. \acp{rb} with strong interference may indicate a close neighbouring cell currently utilising them, or even multiple interfering cells. Low or zero interference \acp{rb} would obviously be very attractive to a \ac{ms}.
The interference power domain is divided into ``Low,'' ``Medium,'' and ``High'' values\footnotemark[\value{footnote}], to categorise \acp{rb} by the amount of interference they suffer. 

\item The \textbf{fast fading component} for each \ac{rb} may not always be readily available, however can become accessible via sounding or pilot/data transmission. 
Users' frequency selective fading profiles extend over the whole available bandwidth, and hence certain \acp{rb} are more suitable to an \ac{ms} than others; or than to other \acp{ms}.
The fast fading domain is split into ``Deep,'' ``Average,'' and ``Peak'' values, centred around the mean fading level~1. In general, \acp{ms} should avoid \acp{rb} with ``Deep'' fades and try to acquire \acp{rb} with ``Peak'' fading values.
\end{itemize}
A graphical representation of the input variables and their ``fuzzification'' is shown in Fig.~\ref{fig:fuzzy_logic_system}.

\subsubsection{Fuzzy System}

The fuzzy logic system is responsible for determining the allocatability of each \ac{rb} in the cell, and the corresponding transmit powers. This is performed in three stages, as can be seen in Fig.~\ref{fig:fuzzy_logic_system}. First, the fuzzified values of the inputs (see. Fig.~\ref{fig:fuzzy_logic_system}) are fed into the \textbf{rule evaluation} stage, where these are combined to determine the ``scores'' of the membership functions of the outputs. These \textbf{rules} are defined in Table~\ref{tab:rules}.
\begin{table}[htb]
\centering
\caption{Fuzzy Rules}
\footnotesize
\begin{tabular}{|c|c|ccccc|ccc|}
\hline
&\textbf{Comb.} & \textbf{Des. Rate} & \textbf{Signal} & \textbf{Interference} & \textbf{Fading} & \textbf{SINR} & \textbf{RB Alloc.} & \textbf{Power} & \textbf{Modulation} \\
\thickhline
1&AND & - & \emph{not} Low & Low & - & - & Yes & Half & - \\
2&AND & Low & \emph{not} Low & Med & Deep & & Yes & Max. &- \\
3&AND & \emph{not} Low & - & High & - & - & No &  - &- \\
4&AND & Low-Med & \emph{not} Low & Med & \emph{not} Deep & - & Yes & Max. &- \\
5&AND & Med-High & \emph{not} Low & Med & Peak & - & Yes & Max. &- \\
6&OR & - & - & High & Deep & - & No & - & - \\
7&AND & - & High & - & \emph{not} Deep &  - & Yes & Half & - \\
8&AND & - & Low & \emph{not} Low & - & - & No & - & -  \\
9&AND & Med-High & High & Med & Peak & - & Yes & Half & - \\
\hline
10& - & - & - & - & - & MuchWorse & - & - & Reduce3 \\
11& - & - & - & - & - & Marg.Worse & - & - & Reduce2 \\
12& - & - & - & - & - & Worse & - & - & Reduce1 \\
13& - & - & - & - & - & Adequate & - & - & NoChange \\
14& - & - & - & - & - & Better & - & - & Increase1 \\
15& - & - & - & - & - & Marg.Better & - & - & Increase2 \\
16& - & - & - & - & - & MuchBetter & - & - & Increase3 \\
\hline
\end{tabular}
\label{tab:rules}
\end{table}
Most of these rules are self-explanatory. In essence, they are intuitive guidelines as to why a specific \ac{rb} should be assigned to the \ac{ms} or not, \eg allocating an \ac{rb} that is receiving high interference (3. and 6.) is not beneficial except in certain cases; or allocating a medium-interference \ac{rb} should not be done if the required rate is too high or the signal level is too low (4. and 5.). Finally, almost any \ac{rb} with low interference can be allocated and be transmitted on with half power to achieve its rate (1.).

After this, in the \textbf{rule output aggregation} stage, the results of all rules are combined for each \ac{rb} to yield a fuzzy set representing \emph{how much} an \ac{rb} \emph{should or should not} be allocated, and \emph{how much} it \emph{should or should not} transmit at half power (\ie if the majority of the rules yield ``Yes'' for \ac{rb} allocation, then the \ac{rb} \emph{should} be allocated \emph{more} than it \emph{should not} be).

Finally, in the \textbf{defuzzification} stage, the \emph{centre of gravity} (which is calculated using the integral-quotient in the Defuzzification box in Fig.~\ref{fig:fuzzy_logic_system}) of the fuzzy set of each output is calculated to give a ``score'' for each \ac{rb}. In essence, this stage determines finally the \ac{rb} allocation (Yes/No) and the \ac{rb} transmit power (Half/Max.), \eg an \ac{rb} allocation score of 0.25 indicates a ``Yes,'' and an \ac{rb} transmit power score of 0.6 recommends maximum power transmission. Clearly, an \ac{rb} with an allocation score of 0.1 is more allocatable than one with a score of 0.4.

\subsubsection{Outputs}

Finally, the output variables of the fuzzy logic system are:
\begin{itemize}
\item The \textbf{\ac{rb} allocation} of the \ac{ms}. The allocatability of each \ac{rb} is calculated by fuzzy logic depending on the inputs. In the end, the \ac{bs} assigns the required number of \acp{rb} to the \acp{ms}, choosing those that are most suitable for each. The lower the score, the better.
\item The \textbf{transmit powers} of the \acp{rb} assigned to the \ac{ms}. Each \ac{rb} can transmit with either half or full (\ie maximum) power, depending on the inputs. For example, an \ac{rb} with low interference may transmit at half power, whereas if the \ac{ms}'s desired signal is low or the fading on that \ac{rb} is deep, full power should be utilised.
\end{itemize}

\subsection{SINR-dependent Link Adaptation}

\begin{figure}[htb]
\centering
\includegraphics[width=.7\textwidth]{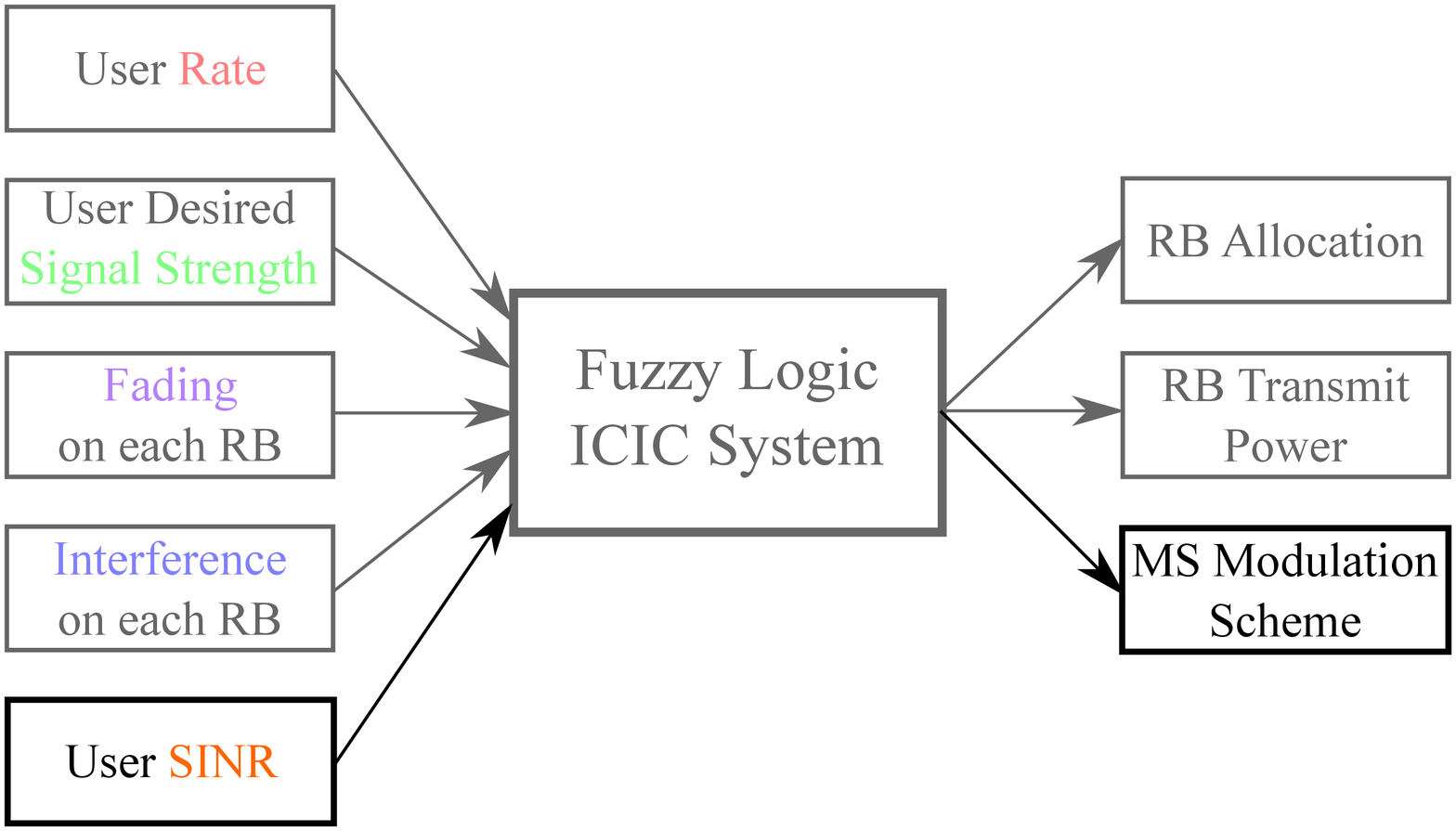}
\caption{Simplified graphical representation of our autonomous resource and power allocation technique with the opportunity for \ac{la}.}
\label{fig:sinr_simp_diagram}
\end{figure}

In general, a wireless channel can change quite rapidly given alterations to its immediate environment, and hence there may be situations where a \ac{ms}'s desired link quality is much better/worse than necessary for its \ac{mcs}. Alternatively, the scenario may arise when the \ac{bs}/\ac{ms} receives high interference from a nearby transmitter, and hence the user's \ac{sinr}~may fall below its target. Therefore, it is imperative that a \ac{ms} can modify its \ac{mcs} depending on the channel conditions. In Fig.~\ref{fig:sinr_simp_diagram}, such an ability is added to the fuzzy logic \ac{icic} system.

Since the success/failure of transmission on a given \ac{rb} is mainly dependent on the \ac{sinr} achieved on it, the \ac{ms} \ac{sinr}\footnote{One might argue that given a user's signal strength and \ac{rb} interference information, that a separate \ac{sinr} input is unnecessary. However, because the \ac{ms} can only receive interference information from other users transmitting on specific \acp{rb}, it is not guaranteed that it receives interfering signals on all \acp{rb}. Furthermore, the desired signal is also only measured on the allocated \acp{rb}, so a standard measure of the average \ac{sinr} is the most precise description of an \ac{ms}'s overall transmission conditions.} is utilised to directly modify the \ac{ms}'s \ac{mcs}: this is called \emph{\ac{la}}. More specifically, the difference between the user's achieved average \ac{sinr} $\bar{\gamma_u}$ and its target $\gamma_u^*$
\begin{equation}
\Delta_{\gamma}= \bar{\gamma_u}-\gamma_u^*\,,
\end{equation}
is utilised. The membership functions for the \ac{sinr} input and \ac{mcs} output are shown in Fig.~\ref{fig:fuzzy_logic_system}. It should be mentioned that only $\Delta_{\gamma}$ is used in the \ac{la} procedure, such that
\begin{itemize}
 \item if $\Delta_\gamma{>}3\,$dB the input is ``Better'', and the \ac{ms} modulation and coding order is ``Increased by 1;''
 \item if $\Delta_\gamma{>}5\,$dB the input is ``Marginally Better'', and the \ac{ms} modulation and coding order is ``Increased by 2;''
 \item if $\Delta_\gamma{>}7\,$dB the input is ``Much Better'', and the \ac{ms} modulation and coding order is ``Increased by 3;''
 \item if $\Delta_\gamma{<}{-}3\,$dB the input is ``Worse'', and the modulation and coding order is ``Reduced by 1;''
 \item if $\Delta_\gamma{<}{-}5\,$dB the input is ``Marginally Worse'', and the modulation and coding order is ``Reduced by 2;''
 \item if $\Delta_\gamma{<}{-}7\,$dB the input is ``Much Worse'', and the modulation and coding order is ``Reduced by 3;'' or lastly
 \item if $-3{<}\Delta_\gamma{<}3\,$dB the input is ``Adequate'', and the modulation and coding order undergoes ``No Change.''
\end{itemize}
These rules are shown in Table~\ref{tab:rules}. Through this procedure, a user may fit its \ac{mcs} to its transmission environment, and hence more easily achieve its target rate. Moreover, the average \ac{sinr} $\bar{\gamma_u}$ is considered to prevent a \ac{ms} from ``ping-pong''-ing between \acp{mcs}, which may severely complicate the scheduling procedure.

\subsection{Scheduling}
\label{sec:fuzzy_sched}

Given the common assumption in femto-cell networks that only a single \ac{ms} is present per cell, 
this user can be allocated the \acp{rb} with the best scores (as determined by the fuzzy logic system). In the reverse link, the contiguity constraint (specific to \ac{lte}) is fulfilled by allocating the required number of consecutive \acp{rb} with the least sum-score. With each \ac{fbs} allocating the most suitable \acp{rb} in their cell, a natural frequency reuse will result. More specifically, it can be shown that neighbouring \acp{fbs} will allocate orthogonal 
sets of \acp{rb}, whereas femto-cells further from each other (\ie less interfering) may assign the same \acp{rb} without excessive interference.

There are, however, many possibilities to perform resource allocation in the presence of multiple users. For instance, in the forward link an \ac{fbs} may simply assign \acp{rb} in the ascending order of scores calculated for all \acp{ms}. This is clearly a greedy approach, and may not be optimal in cases where \acp{ms} have vastly different channel conditions (not usually the case in femto-cells, but possible). Another possibility, then, for resource allocation may be a \ac{pfs}, where the \ac{rb} scores for each user are scaled by the ratio of achieved and desired rates. Here, an \ac{ms} that strongly underachieved its rate would be allocated \acp{rb} before an \ac{ms} that was closer to its target. Lastly, a ``priority'' scheduler may be utilised to give precedence to users with higher required rates/modulation orders, to more likely fulfil their QoS requirements.

\subsection{Signal Statistics}

In Fig.~\ref{fig:fuzzy_logic_system}, the membership functions of the desired and interfering signal inputs are determined via analysis of the signal statistics in the deployment environment. While these can be determined experimentally, we analytically derive here these statistics such that they can be expanded to other scenarios. Thus, we know the power of any received signal $P_r$ is calculated as
\begin{align}
P_r &= P_t G \notag\\
P_{r,\rm dB} &= P_{t,\rm dB} + G_{\rm dB} = P_{t,\rm dB} - L_{\rm dB}\label{eq:p_r}
\end{align}
where $L_{\rm dB}{=}L_{d,\rm dB}+X_\sigma$ is the signal path loss, and $L_{d}$ and $X_\sigma$ are described in Section~\ref{sec:chanmod}. Hence, the \ac{pdf} of $P_r$ (in dB) is given by
\begin{equation}
f_{P_r,\rm dB}(\varrho) = f_{P_t,\rm dB}(\theta) \circledast f_{L,\rm dB}(-l;D)\,.
\end{equation}
where $\circledast$ denotes the convolution operator. And since
\begin{equation}
f_{L,\rm dB}(l;D) = f_{L_d,\rm dB}(l) \circledast f_{X_\sigma,\rm dB}(x)\,,
\end{equation}
by finding $f_{L_d,\rm dB}(l)$ and $f_{P_t,\rm dB}(\theta)$, $f_{P_r,\rm dB}(\varrho)$ is derived for both desired and interfering signals.

\begin{figure}[htb]
\centering
\subfigure[Path Loss and Transmit Power PDFs]{\label{sfig:pl_tp}
\includegraphics[width=.48\columnwidth]{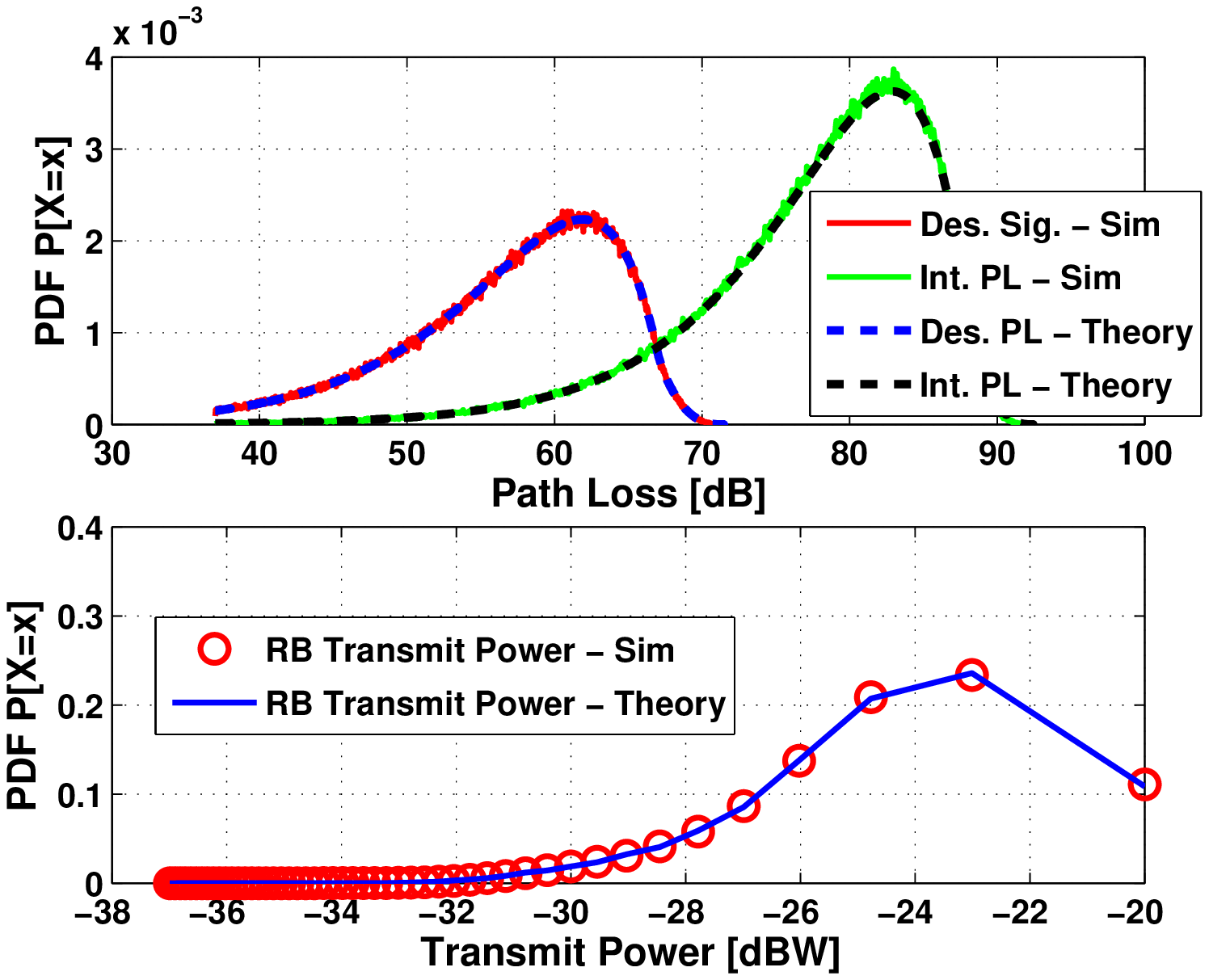}}
\hfill\subfigure[Signal Energy CDFs]{\label{sfig:sigcdfs}
\includegraphics[width=.48\columnwidth]{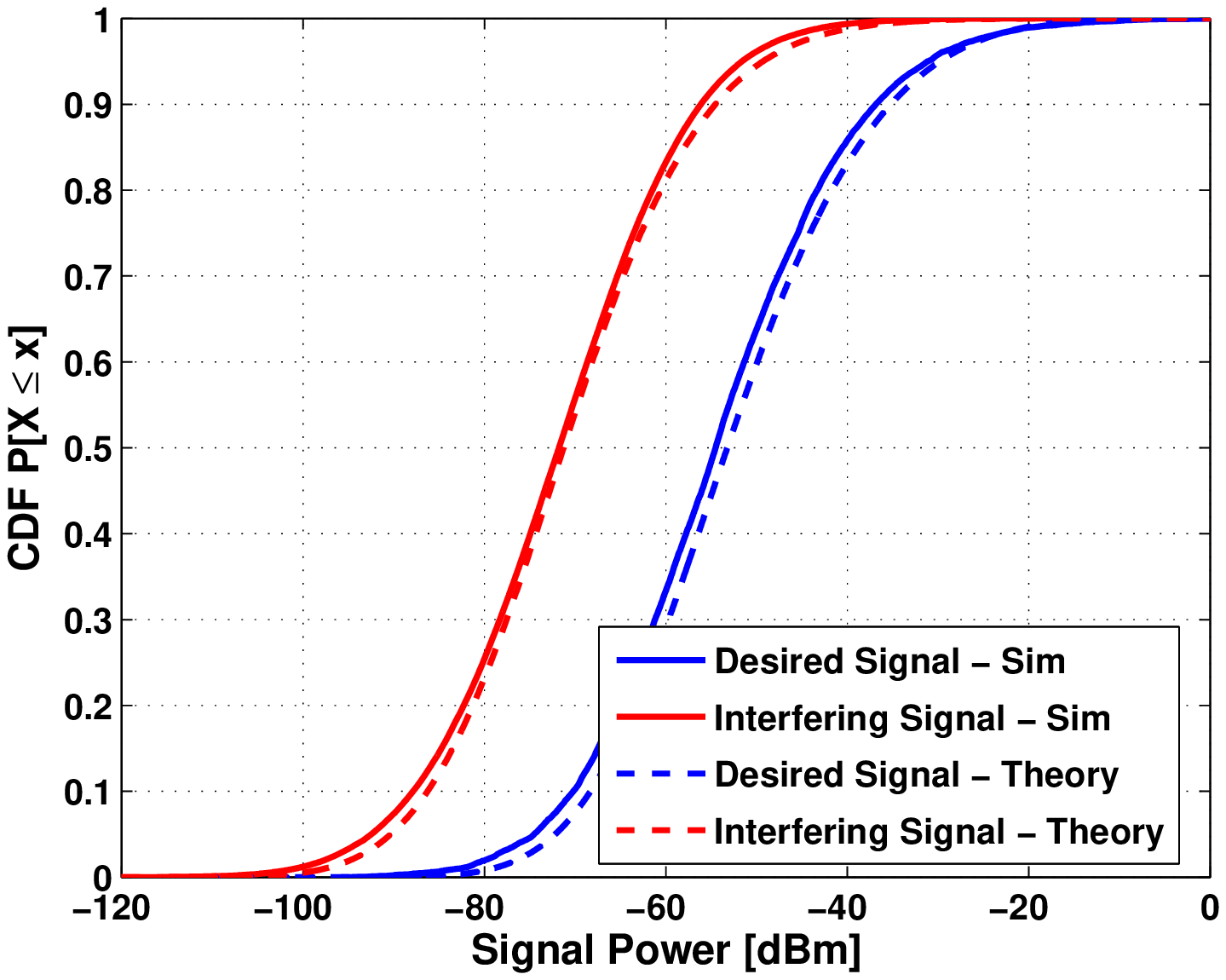}}
\caption{Comparison of derived theoretical desired and interfering signal \acp{pdf} and \acp{cdf} to Monte Carlo simulation results, considering lognormal shadowing.}
\label{fig:sig_stats}
\end{figure}

Due to the random nature of the \ac{bs} and \ac{ms} positions, the first step in analysing the signal \acp{pdf} is estimating the distribution of the path losses between transmitter (whether it is desired or interfering) and receiver. From~\eqref{eq:pathlossfunc} it is clear that the path loss $l$ is proportional to the Tx-Rx distance $d$, and the inverse relationship is given by
\begin{equation}
\rho(l) = d = 10^{\nicefrac{(l-\alpha)}{\beta}}\,.
\end{equation}
Hence, the distance dependent loss \ac{pdf} $f_{L_d,\rm dB}(l;D)$ is calculated by
\begin{align}
f_{L_d,\rm dB}(l;D) &= \left\lvert \frac{{\rm d}\rho(l)}{{\rm d}l}\right\rvert f_d(\rho(l);D)\\
\left\lvert \frac{{\rm d}\rho(l)}{{\rm d}l}\right\rvert&=\frac{\ln10}{\beta}10^{\nicefrac{l-\alpha}{\beta}} =\frac{\ln10}{\beta}\rho(l)\notag\\
f_{L_d,\rm dB}(l;D) &= \frac{\ln10}{\beta}\rho(l) f_d(\rho(l);D)\,,
\label{eq:pdf_l_calc}
\end{align}
where $f_d(\rho(l);D)$ is the \ac{pdf} of the Tx-Rx distance parametrised by the dimension $D$. This \ac{pdf} is given in~\cite{t0401} by
\begin{align}
&f_d(d;D)= \notag\\
&= \left\{
\begin{array}{ll}
\!\!2\frac{d}{D}\left(\left(\frac{d}{D}\right)^2-4\frac{d}{D}+\pi\right) & 0\leq d\leq D \\
\!\!2\frac{d}{D}\left[4\sqrt{\left(\frac{d}{D}\right)^2-1} - \left(\left(\frac{d}{D}\right)^2+2-\pi\right)-4\tan^{-1}\left(\sqrt{\left(\frac{d}{D}\right)^2-1}\right)\right] & D<d\leq\sqrt{2}D
\end{array}
\right.\!\!\!.
\label{eq:pdf_d}
\end{align}
Thus, by evaluating~\eqref{eq:pdf_d} as in~\eqref{eq:pdf_l_calc}, the distance-dependent path loss \ac{pdf} $f_{L_d,\rm dB}(l;D)$ becomes
\begin{align}
&f_{L_d,\rm dB}(l;D) =\notag\\
&= \frac{\ln10}{\beta}\rho(l)\left\{
\begin{array}{ll}
\!\!2\delta(l)\left(\delta(l)^2-4\delta(l)+\pi\right) & \alpha\leq l\leq L(D) \\
\!\!2\delta(l)\left[4\sqrt{\delta(l)^2{-}1} {-} \left(\delta(l)^2{+}2{-}\pi\right){-}4\tan^{{-}1}\left(\sqrt{\delta(l)^2-1}\right)\right] & L(D){<}l{\leq}L\left(\sqrt{2}D\right)
\end{array}
\right.\!\!\!\!,
\label{eq:pdf_l}
\end{align}
where $\delta(l){=}\nicefrac{\rho(l)}{D}$. This \ac{pdf} can be seen for both the desired signal ($D{=}10$m) and the interfering signal ($D{=}50$m, as interferer and receiver could be located in any two apartments in the scenario) in Fig.~\ref{sfig:pl_tp}. Monte Carlo simulations that randomly place two nodes within the given dimensions $D{\times}D$, and calculate the resulting path loss, verify that the \ac{pdf} given in~\eqref{eq:pdf_l} is indeed correct. 

Referring back to~\eqref{eq:p_r}, we have accurately described the path loss $L_{\rm dB}$, and must now find the distribution of the \ac{rb} transmit powers $P_t$. In our model, each \ac{ms} transmits with a maximum \emph{total} power $P_{\max}$ that is spread evenly over all \acp{rb} assigned to it. The number of \acp{rb} $n_{\rm RB}$ an \ac{ms} is assigned is directly dependent on the required rate $C^*_u$ of the user, thus $P_t$ is defined by
\begin{align}
P_t &= \frac{P_{\max}}{n_{\rm RB}} \qquad \mbox{where  } n_{\rm RB}=\left\lceil \frac{C^*}{k_{\rm sc}s_{\rm sc}} \right \rceil\notag\\
&= \frac{P_{\max}k_{\rm sc}s_{\rm sc}}{C^*} = \frac{A}{C^*}\label{eq:p_t}\,.
\end{align}
Here, the ceiling operation is removed for ease of derivation, however without loss of generality. Therefore, it is clear from~\eqref{eq:p_t} that $P_t$ is inversely proportional to the rate $r$, which in our scenario is a random variable with distribution $f_{C^*}(r)$. Hence, the \ac{cdf} of the transmit power $F_{P_t}(p)$ is given by
\begin{align}
F_{P_t}(p) &= {\rm\mathbf{P}}\left[ P_t\leq p\right]= {\rm\mathbf{P}}\left[ \frac{A}{r}\leq p\right]= {\rm\mathbf{P}}\left[ \frac{A}{p}\leq r\right]\notag\\
&= 1-{\rm\mathbf{P}}\left[ r\leq \frac{A}{p}\right] = 1-F_{C^*}\left(\frac{A}{p}\right)\,,\notag
\end{align}
where $F_{C^*}(r)$ is the \ac{cdf} of user desired rates, and therefore the \ac{pdf} of the \ac{ms} transmit power $f_{P_t}(p)$ is given by
\begin{align}
f_{P_t}(p) &= \frac{{\rm d}F_{P_t}(p)}{{\rm d}p} = \frac{A}{p^2}f_{C^*}\left(\frac{A}{p} \right)\label{eq:pdf_p_t}
\end{align}
The general expression is given in~\eqref{eq:pdf_p_t} for any rate \ac{pdf} $f_{C^*}(r)$. 
Now, we need to perform a change of variable transform to determine the \ac{pdf} of the transmit power in dB (refer to~\eqref{eq:p_r})
\begin{equation}
\theta = P_{t,\rm dB} = 10\log_{10}(P_t)\,,
\end{equation}
and the inverse is given by
\begin{equation}
\varphi(\theta) = p = 10^{\nicefrac{\theta}{10}}\,.
\end{equation}
Thus, the \ac{pdf} of \ac{ms} transmit power $f_{P_t,\rm dB}(\theta)$ is calculated by
\begin{align}
f_{P_t,\rm dB}(\theta) &= \left\lvert \frac{{\rm d}\varphi(\theta)}{{\rm d}\theta}\right\rvert f_{P,\rm dB}(\varphi(\theta))\\
\left\lvert \frac{{\rm d}\varphi(\theta)}{{\rm d}\theta}\right\rvert&=\frac{\ln10}{10}10^{\nicefrac{\theta}{10}} =\frac{\ln10}{10}\varphi(\theta)\,,\notag
\end{align}
hence
\begin{align}
f_{P_t,\rm dB}(\theta) &= \frac{\ln10}{10}\varphi(\theta) f_{P_t}(\varphi(\theta))\notag\\
&= \frac{\ln10}{10}\frac{A}{\varphi(\theta)}f_{C^*}\left(\frac{A}{\varphi(\theta)} \right)\label{eq:pdf_p_t_dB}
\end{align}
where~\eqref{eq:pdf_p_t_dB} is the general expression for any rate distribution. Thus, the \ac{pdf} of user transmit power has been derived, however under the assumption of transmission of a single bit per channel use. This is, of course, not a realistic assumption, and in our scenario we consider a user's ability to send with various \acp{mcs} (see Table~\ref{tab:amc}). Clearly, the \ac{mcs} affects the number of \acp{rb} required by an \ac{ms}, and thus also the \ac{ms} transmit power. This is shown in~\eqref{eq:p_t_m}
\begin{align}
P_t &= \frac{P_{\max}}{n_{\rm RB}} \qquad \mbox{where } n_{\rm RB}=\left\lceil \frac{{C^*}}{k_{\rm sc}s_{\rm sc}\varepsilon_{\rm s}} \right \rceil\notag\\
&= \frac{P_{\max}k_{\rm sc}s_{\rm sc}\varepsilon_{\rm s}}{{C^*}} = \frac{A\varepsilon_{\rm s}}{{C^*}}\label{eq:p_t_m}\,.
\end{align}
Further, we assume each user is uniformly distributed a \ac{mcs}\footnote{This would be independent of its signal quality. This is not the best assumption, admittedly, however the reason is to further randomise the user requirements, and hence the necessary \ac{rb} allocations. Through this, the allocation problem becomes more challenging for \ac{icic} techniques, including our own.},  
hence by replacing~\eqref{eq:p_t} with~\eqref{eq:p_t_m} and performing the same \ac{cdf} transformation, the transmit power \acp{pdf} (\ie $f_{P_t}(p)$ and $f_{P_t,\rm dB}(\theta)$) are modified correspondingly as
\begin{align}
f_{P_t}(p) &\rightarrow \frac{1}{4} \sum_{m=0}^{15}\frac{A\varepsilon_{\rm s}(m)}{p^2}f_{C^*}\left(\frac{A\varepsilon_{\rm s}(m)}{p} \right)\,,\notag\\
f_{P_t,\rm dB}(\theta) &\rightarrow \frac{\ln10}{40} \sum_{m=0}^{15}\frac{A\varepsilon_{\rm s}(m)}{\varphi(\theta)}f_{C^*}\left(\frac{A\varepsilon_{\rm s}(m)}{\varphi(\theta)} \right)\label{eq:pdf_p_t_m_dB}
\end{align}
where $m$ is the CQI index in Table~\ref{tab:amc}, and again,~\eqref{eq:pdf_p_t_m_dB} is the general expression 
for any user rate distribution. Now, if we revisit that $n_{\rm RB}{=}\left\lceil \frac{{C^*}}{k_{\rm sc}s_{\rm sc}\varepsilon_{\rm s}}\right\rceil$, it is clear that only integer number of \acp{rb} can be assigned to each \ac{ms}, and thus each user can only assume a transmit power from a discrete set of $P_t{=}\frac{P_{\max}}{n_{\rm RB}}$
\begin{equation}
P_t \in \left\{\frac{P_{\max}}{1}, \frac{P_{\max}}{2},\cdots,\frac{P_{\max}}{M}\right\}\label{eq:p_t_disc}\,,
\end{equation}
where $M$ denotes the total number of \acp{rb} available in each cell. Thus, $f_{P_t}(p)$ is evaluated at the powers in~\eqref{eq:p_t_disc}, as are the histogram bins in the Monte Carlo simulation, the results of which are presented in Fig.~\ref{sfig:pl_tp} for $C^*{\sim}{\rm Rayl}\left(\bar{C}\right)$, where $\bar{C}$ is the average rate. The close match of theoretical and empirical results confirms that the derivation for $f_{P_t,\rm dB}(\theta)$ is indeed accurate.

Thus, we have now found accurate and precise analytical models for the distributions of the path losses and transmit powers, which are directly dependent on the network topology of the investigated scenario. From~\eqref{eq:p_r} it is clear that
\begin{equation}
f_{P_r,\rm dB}(\varrho) = f_{P_t,\rm dB}(\theta) \circledast f_{L,\rm dB}(-l;D)\,.
\end{equation}
Hence, the desired and interfering signal \acp{pdf} are given in~\eqref{eq:pdf_s} and~\eqref{eq:pdf_i}, respectively,
\begin{align}
f_{S,\rm dB}(s) &= f_{P_t,\rm dB}(\theta) \circledast f_{L,\rm dB}(-l;D{=}10)\label{eq:pdf_s}\\
f_{I,\rm dB}(i) &= f_{P_t,\rm dB}(\theta) \circledast f_{L,\rm dB}(-l;D{=}50)\,.\label{eq:pdf_i}
\end{align}
In Fig.~\ref{sfig:sigcdfs} a comparison to simulation results is drawn, where it is evident that the theoretical \acp{cdf} are slightly shifted from their experimental counterparts. The general shape (\ie variance) of the \acp{cdf} is accurate, and while there is a minor shift (1-2$\,$dB) between simulation and theory, we feel that this difference is within the numerical margin of error, and thus acceptable.

It is clear that the signal strength \acp{pdf} are mainly dependent on the distance between transmitter and receiver, and the transmit power. Therefore, extending fuzzy logic \ac{icic} to other scenarios is straightforward, as simply the distance \ac{pdf} $f_d(d;D)$ must be modified to fit the new environment, and the statistics can be found. Hence, not only have the desired and interfering signals been derived for the femto-cell scenario, they are easily modified to other environments, thus expanding the applicability of fuzzy logic \ac{icic} to virtually any wireless network.

\section{Optimality of Fuzzy Logic ICIC}
\label{sec:optim}

Due to the heuristic nature and non-linearity of fuzzy logic, it is very difficult to perform a comprehensive theoretical analysis of the system performance of fuzzy logic \ac{icic}. Therefore, to analyse the optimality of our technique, we perform an experimental comparison between fuzzy logic \ac{icic} and two well-known forms of resource allocation. 
We demonstrate that fuzzy logic \ac{icic} provides close-to-optimal throughput and coverage at significantly reduced complexity.

\subsection{System Optimisation}

The most obvious choice for performance comparison is that of posing the resource allocation as a system-wide optimisation problem. Since fuzzy logic is autonomous and, more importantly, distributed it \emph{should}, on average, be suboptimal in terms of overall system performance. The optimal \ac{rb} allocation of the system can be achieved by solving the problem posed in~\eqref{eq:opt_prob}, and thus the aim of fuzzy logic is to as closely as possible approach the result of this problem. Given the definition for user throughput~\eqref{eq:C_u} and system sum throughput~\eqref{eq:c_sys_def}, we solve
\begin{alignat}{3}
\max \quad& C_{\rm sys}=\sum_u C_u &\quad& u{=}1,\,2,\,\dots\,,\,n_{\rm usr}\,.\label{eq:opt_prob}\\
\mbox{s.t.} \quad& \sum^M_{j=1} \vect{1}_{P_{u,j}>0} = n_u^{\rm RB} &\quad&\forall u \label{eq:nrb_cons}\tag{\ref{eq:opt_prob}a}\\
& \sum^M_{j=1} P_{u,j} \leq P_{\max}&\quad&\forall u\label{eq:pmax_cons}\tag{\ref{eq:opt_prob}b}\\
& P_{u,j}\geq0 &\quad&\forall u,j\label{eq:p_cons}\tag{\ref{eq:opt_prob}c}
\end{alignat}
in order to determine the maximum rate achievable in a given scenario. In the constructed \ac{minlp}~\cite{book:b9901} problem, \eqref{eq:pmax_cons} and~\eqref{eq:p_cons} describe the restrictions on transmit power allocation at each \ac{ms}: the sum of the allocated powers on all \acp{rb} cannot exceed $P_{\max}$, and the individual powers must be non-negative, respectively. The constraint~\eqref{eq:nrb_cons} limits the number of transmitting \acp{rb} at a single \ac{ms} to the $n_u^{\rm RB}$ the user needs to achieve its desired rate. This is necessary as since the objective is sum-rate-maximisation, the best solution is generally transmission on most, if not all \acp{rb}. However, since fuzzy logic \ac{icic} only aims to satisfy user requirements\footnote{It should be mentioned that a minimum rate constraint was originally considered. However, if a single \ac{ms} cannot achieve its target rate, then no solution can be found by the problem, and hence this constraint was removed.}, this would be an unfair comparison; hence the constraint~\eqref{eq:nrb_cons}. 

\subsection{Greedy Heuristic}

While the comparison to the system-wide optimisation problem will demonstrate the optimality of fuzzy logic \ac{icic}, it is important to note that we are comparing a centralised and a distributed approach. Therefore, we implement a commonly utilised distributed allocation technique, which ``greedily'' allocates the best \acp{rb} to the \ac{ms}(s) in the cell~\cite{lmssn0901}. Here, the potential \ac{sinr} achievable on each \ac{rb} is calculated using prior interference, signal, and transmit power information; and then the \acp{rb} with the strongest \acp{sinr} will be allocated to the user.
\begin{equation}
\begin{split}
\mbox{Given:} \quad& P_u=\nicefrac{P_{\max}}{n_u^{\rm RB}},\,\,I_{u}^m,\,\, G_{u,v_u}^m \quad m{=}1,\,2,\,\dots\,,\,M\,,\\
\mbox{Find:} \quad & \gamma_{u}^m=\frac{P_u G_{u,v_u}^m}{I_{u}^m+\eta} \qquad \forall m\,.
\end{split}\label{eq:greed_heur}
\end{equation}
In~\eqref{eq:greed_heur}, the same information is available as for fuzzy logic, and a greedy approach is utilised to allocate the \acp{rb}. This technique should maximise the throughput in each cell, however it does not take a system view as in~\eqref{eq:opt_prob}, and hence will be suboptimal in terms of network throughput.

Therefore, we argue that the fuzzy logic \ac{icic} comparison to this greedy heuristic will show the optimality of fuzzy logic on an individual cell basis, whereas the comparison to the optimisation problem will show the optimality achieved at the network level.

\subsection{Results Comparison}

\begin{figure}[htb]
\centering
\subfigure[Throughput]{\includegraphics[width=.49\columnwidth]{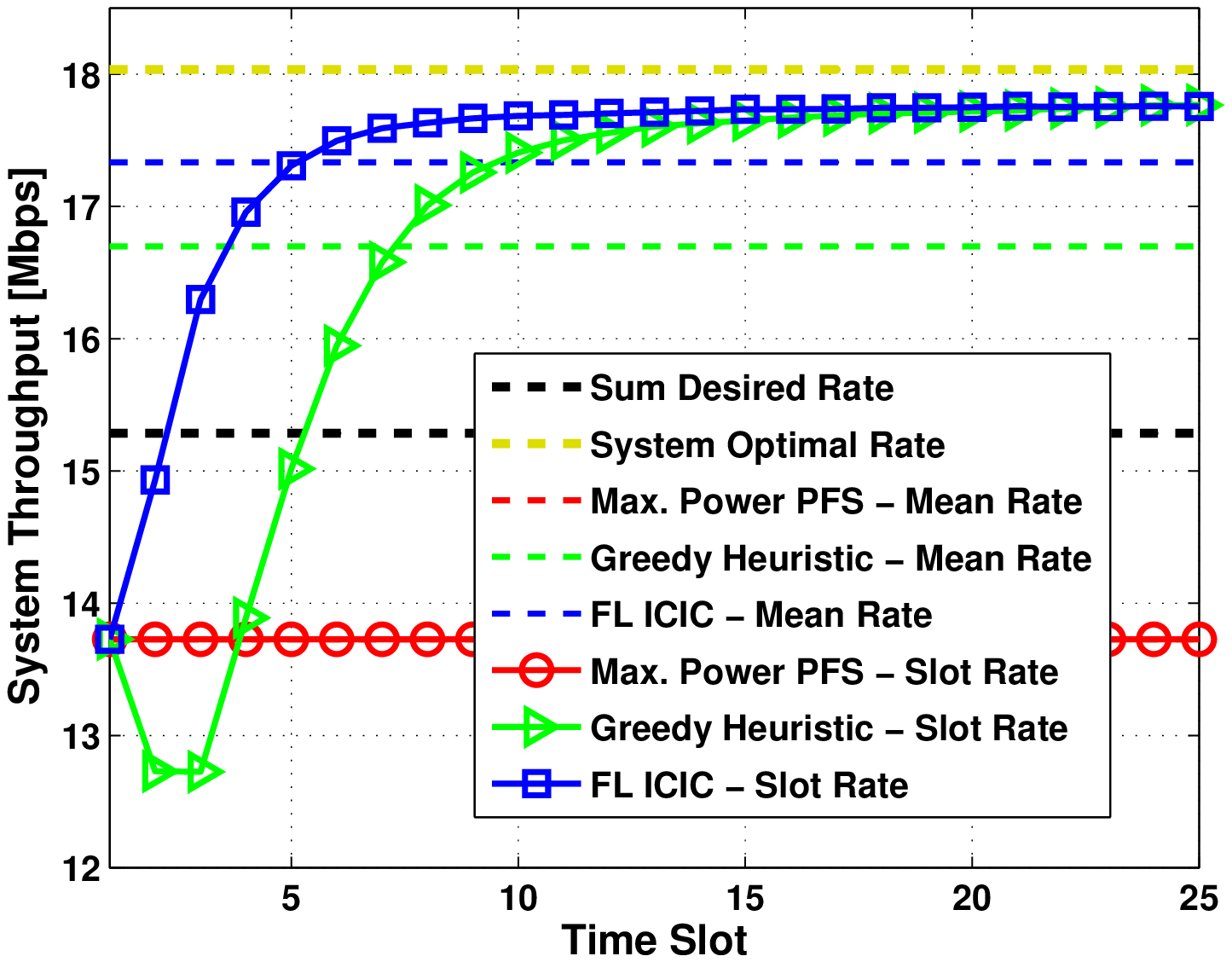}}
\subfigure[Availability]{\includegraphics[width=.49\columnwidth]{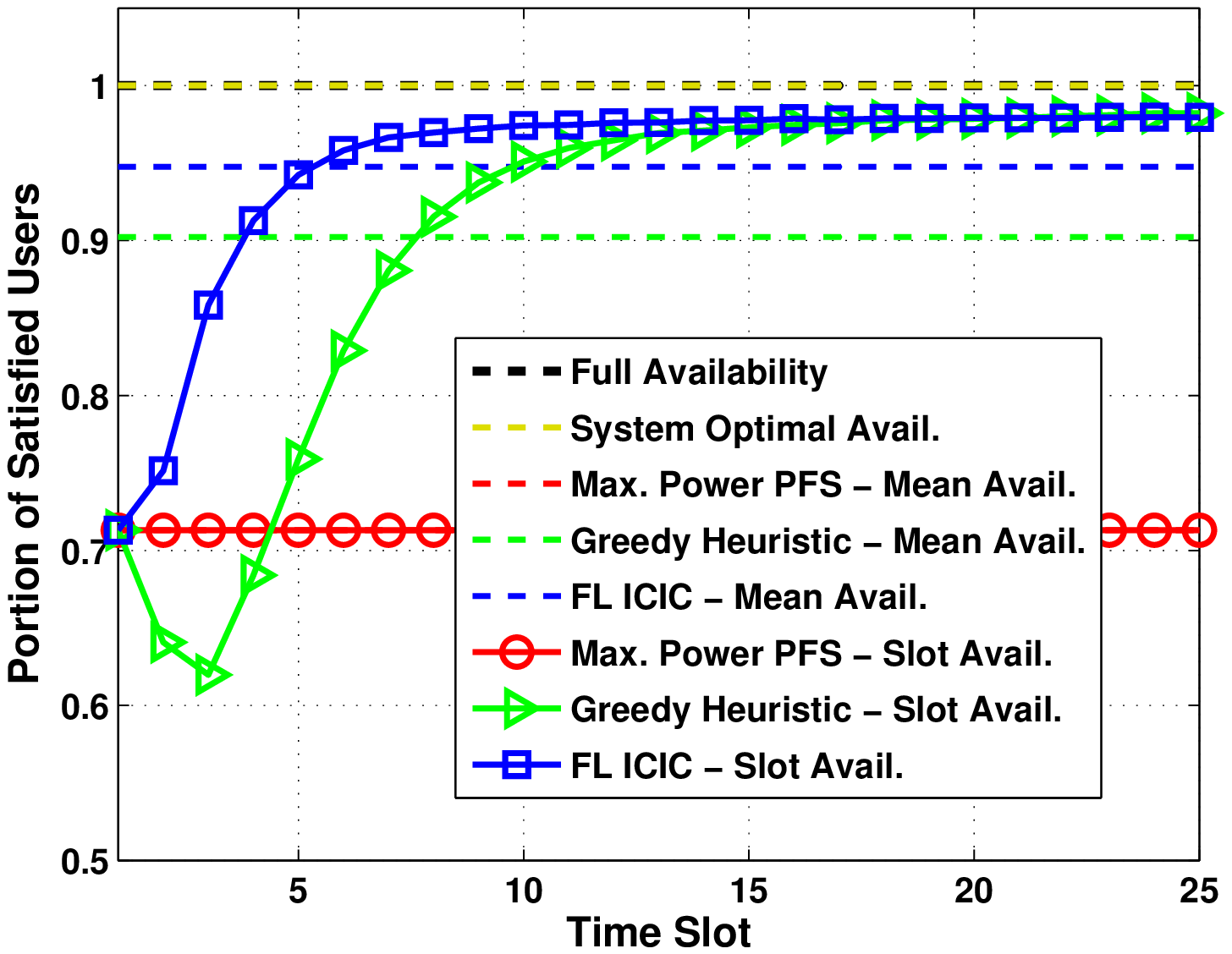}}
\caption{System performance comparison of fuzzy logic \ac{icic}, the system-wide optimal solution, and the proposed greedy heuristic.}
\label{fig:perf_comp}
\end{figure}

To compare the performance of these three methods, a Monte Carlo simulation is run utilising the ${5{\times}5}$ apartment grid model described in Section~\ref{sec:scenmodel}, with $\tilde{\mu}(u){=}1$, and $\bar{C}{=}1.25\,$Mbps. We utilise standard fuzzy logic \ac{icic} without \ac{la}, as neither the optimisation technique nor the greedy heuristic employ \ac{la}. Fig.~\ref{fig:perf_comp} shows the throughput and availability results for this scenario, where it is evident that the system-optimum solution cannot be reached by the distributed techniques.
However, fuzzy logic is able to perform, on average, within 4\% of the optimum throughput performance, and in fact the difference after 20 time slots (\ie two \ac{lte} frames) is less than 2\%. Furthermore, it is clear that the average throughput of fuzzy logic is improved over the greedy heuristic (by 4\%), even though after 15 time slots the performance is similar. This highlights that fuzzy logic \ac{icic} is optimal on a cell-individual basis, however is able to (due to other inputs such as rate requirement and desired signal strength) converge to this optimum much quicker\footnote{The substantial decline in performance by the greedy heuristic in the first time slots results from the lack of interference information. The unused \acp{rb} with ``zero'' interference are allocated in all cells simultaneously, thus causing large outages in these slots. After more accurate statistics have been received, the performance improves as expected.}. On the other hand, the performance difference to the optimum is minute, and therefore fuzzy logic \ac{icic} provides a ``near-optimal'' solution for the network as a whole.

The same trends can be seen for the system availability, where while the optimum is clearly full availability (\ie $\chi{=}1$), fuzzy logic \ac{icic} achieves 98\% coverage, and hence produces almost negligible outage. Furthermore, it is able to reach this availability much faster than the greedy heuristic, indicating that fuzzy logic \ac{icic} employs a balance between system-wide optimisation and cell-individual performance.

\subsection{Complexity}

To conclude our comparison, we analyse the complexities of the three schemes, to highlight the simplicity and efficiency of our fuzzy logic technique. 
In a cell where fuzzy logic \ac{icic} is applied, $K{=}4$ inputs (see Fig.~\ref{fig:simplified_diagram}) are combined at each of $M$ \acp{rb} available at the \ac{fbs}, inducing a complexity of $KM$. Following this, the \acp{rb} are sorted according to their fuzzy score, in order to allocate the most appropriate to the \ac{ms}. Since, in general, sorting algorithms demonstrate $O(N^2)$ complexity, the fuzzy logic complexity within a cell increases to $(KM)^2$. Finally, given a scenario with $n_{\rm usr}$ \acp{ms}, the system complexity of fuzzy logic \ac{icic} is given by
\begin{equation*}
O_{\rm FL}(n_{\rm usr}(KM)^2)\,.
\end{equation*}

The greedy heuristic utilises a similar methodology as fuzzy logic, in that it also computes a ``score'' (in this case the instantaneous \ac{sinr}) for each \ac{rb} and then orders them for allocation. Hence, the evaluation complexity at each \ac{rb} is $KM$ (where in this case $K{=}2$ inputs), the sorting complexity is $(KM)^2$ and the overall complexity is given by
\begin{equation*}
O_{\rm GH}(n_{\rm usr}(KM)^2)\,.
\end{equation*}

\begin{figure}[htb]
\centering
\includegraphics[width=.7\columnwidth]{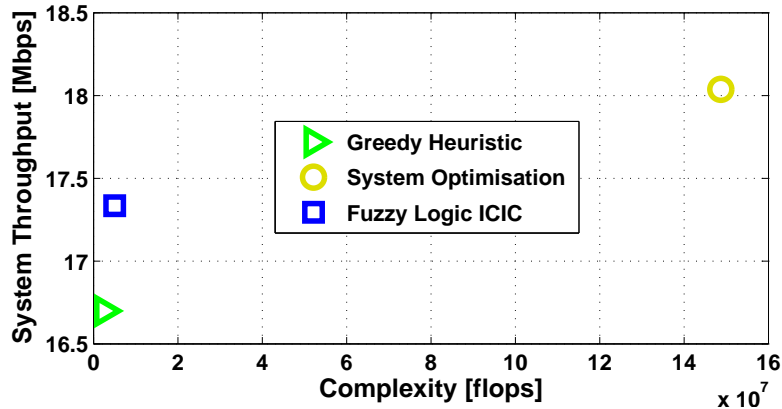}
\caption{System throughput versus required complexity for fuzzy logic \ac{icic}, the system-wide optimal solution, and the proposed greedy heuristic.}
\label{fig:thrpt_comp}\vspace*{-20pt}
\end{figure}

For the optimisation problem~\eqref{eq:opt_prob}, finding the solution complexity is more challenging than for the heuristics, as the problem is considered $\mathcal{NP}$-hard~\cite{book:b9901}. In general, $\mathcal{NP}$-hard problems are only solvable (if possible) in exponential time. 
Here, we want to simultaneously find the resource allocation of $n_{\rm usr}$ \acp{ms}, each wishing to allocate $n^{\rm RB}_u$ of the $M$ \acp{rb} available to it. In the worst-case, an exhaustive search must be performed where all allocation possibilities at the \acp{ms} must be tested. Therefore, we estimate the complexity of~\eqref{eq:opt_prob} as
\begin{equation*}
O_{\rm OP}\left(\prod_{u=1}^{n_{\rm usr}} 
\left(\!\!\!\begin{array}{c}
M \\
n^{\rm RB}_u
\end{array}\!\!\!\right)\right)\,.
\end{equation*}
This is clearly much greater than the complexity of the two heuristics, which is expected. A comparison of the achieved throughputs and required complexities\footnote{Due to the massive complexity of the optimisation technique, the $x$-axis in Fig.~\ref{fig:thrpt_comp} is given in dB-flops (dBf${\triangleq}10\log_{10}\left(O_x(\cdot)\right)$), such that results can be compared.} of the three techniques is shown in Fig.~\ref{fig:thrpt_comp}.

It is evident that, while~\eqref{eq:opt_prob} provides the greatest system throughput, it is substantially (\ie exponentially) more complex than both fuzzy logic and the \ac{sinr} heuristic, which only suffer slightly in terms of achieved throughput. On the other hand, it is clear that fuzzy logic \ac{icic} provides enhanced throughput and coverage for the system compared to the greedy heuristic, even though the complexities are very similar. Hence, we conclude that fuzzy logic provides low-complexity, near-optimal system performance in an autonomous and distributed manner.
\section{Simulation}
\label{sec:simulator}

Monte Carlo simulations are used to provide performance statistics of the system with fuzzy logic \ac{icic} and two benchmarks. The simulator is built following \ac{lte} specifications.

\subsection{Scenario Construction and User Distribution}

A ${5{\times}5}$ apartment grid is considered for the simulation environment with $\tilde{\mu}(u){=}3$ (see Fig.~\ref{fig:apart_scen}), and is constructed as described in Section~\ref{sec:scenmodel}. In order to obtain statistically relevant results, the random effects from \ac{ms}/\ac{bs} placement, lognormal shadowing and frequency selective fading must be removed. Therefore, 2000 scenarios (with minimum three \acp{fbs}) are simulated and the results combined to acquire mean performance statistics of the system.

\subsection{Resource Allocation}

Each \ac{ms} is assigned two transmission requirements: a desired throughput and \ac{mcs}. The desired rate $C_u^*$ of each user is drawn from a random distribution\footnote{The distribution can be dependent on the scenario and traffic/applications (\ie internet, mobile TV, etc.) desired by the users.} with mean $\bar{C}$. Due to this, each \ac{ms}$_u$ will require a different number of \acp{rb} $n^{\rm RB}_u$, and hence the system will function best when strongly interfering \acp{fbs} are assigned orthogonal resources.

The \ac{mcs} is also assigned randomly, with equal probabilities for all available symbol efficiencies. While this is not the most realistic assumption\footnote{When \ac{la} is applied, the user's \ac{mcs} will more accurately reflect its \ac{sinr} conditions. Furthermore, the number of \acp{rb} requested will clearly change dependent on the modulation order selected.}, it has been applied here to further randomise the number of \acp{rb} each \ac{ms} needs to achieve its rate.

Finally, \acp{rb} are allocated individually in each cell by the \ac{fbs}. In the benchmarks, a \ac{pfs} is used for \ac{rb} assignment, which improves the frequency diversity relative to a random allocation. On the other hand, the fuzzy logic \ac{icic} technique autonomously allocates \acp{rb} based on the local information available, in order to optimise the \ac{ms}(s) performance in the cell. For our purposes, the allocation of \acp{rb} to \acp{ms} is performed greedily, as described in Section~\ref{sec:fuzzy_sched}.

\subsection{Time Evolution}

Each run of the Monte Carlo simulation is iterated over ${z{=}25}$ subframes, or, equivalently, $2.5$ \ac{lte} frames, such that long-term \ac{sinr} statistics can be gathered. Due to the random user and \ac{fbs} distribution, plentiful runs with different network generations are considered in order to obtain statistically accurate results. 
At the start of each subframe, the scheduling and allocation of \acp{rb} is reperformed. The \acp{ms} are assumed to be quasi-static for the duration of a run.

The simulation is performed for a constant-traffic model, where each user requests the same number of \acp{rb} in each time slot (\ie subframe). Furthermore, the users are assumed to be static for the duration of a subframe, such that effects due to Doppler spread can be neglected. Perfect synchronisation in time and frequency is assumed, such that intra-cell interference is avoided. The relevant simulation parameters can be found in Table~\ref{tab:sim_param}. 
\begin{table}[htb]
\centering
\caption{Simulation Parameters}
\footnotesize
\begin{tabular}{ll}
\thickhline
Parameter & Value \\
\hline
Apartment width, $W$ & $10\,{\rm m}$ \\
\ac{fbs} probability, ${p_{\rm act}}$ & 0.5 \\
Number of available \acp{rb}, $N_{\rm RB}$ & 50 \\
\ac{rb} bandwidth, $B_{\rm RB}$ & $180\,{\rm kHz}$ \\
Average rate, $\bar{C}$ & $1.25\,$Mbps \\
Subcarriers per \ac{rb}, $k_{\rm sc}$ & 12 \\
Symbol rate per subcarrier, $s_{\rm sc}$ & $15\,{\rm ksps}$ \\
Time slots & 25 \\
\ac{abs} prob., $\Gamma_{\rm \!ABS}$ & 0.1 \\
Spectral noise density, $\eta_0$ & $-174\,{\rm dBm/Hz}$ \\
Total \ac{fbs} transmit power & $10\,{\rm dBm}$ \\
Channel parameters $\alpha,\,\beta$ & 97, 30 \\
Shadowing Std. Dev., $\sigma$ & $10\,{\rm dB}$ \\
Auto-correlation distance & $50\,{\rm m}$ \\
\thickhline
\end{tabular}
\label{tab:sim_param}
\end{table}

\subsection{Benchmarks}

To evaluate the performance of fuzzy logic \ac{icic}, two well-known benchmark systems have been implemented for comparison purposes. These are:
\begin{itemize}
\item \textbf{Maximum Power Transmission}: In the first benchmark, no power allocation is performed, and all \acp{ms} transmit at the maximum power on each \ac{rb}. 
\item \textbf{Random \ac{abs} Transmission}: In the second benchmark, again all links transmit at full power, however, in each time slot a user transmits an \ac{abs} with probability $\Gamma_{\rm \!ABS}$, where for this simulation $\Gamma_{\rm \!ABS}{=}0.1$.
\end{itemize}

\section{Results and Discussion}
\label{sec:results}

From the simulation, the statistics of the system throughput, energy efficiency, availability and fairness are generated for systems employing fuzzy logic \ac{icic} and compared against the two benchmark systems. General simulation parameters are taken from Table~\ref{tab:sim_param} and~\cite{std:3gpp-x2gaap}.

\begin{figure}[htb]
\centering
\subfigure[Throughput]{\includegraphics[width=.49\columnwidth]{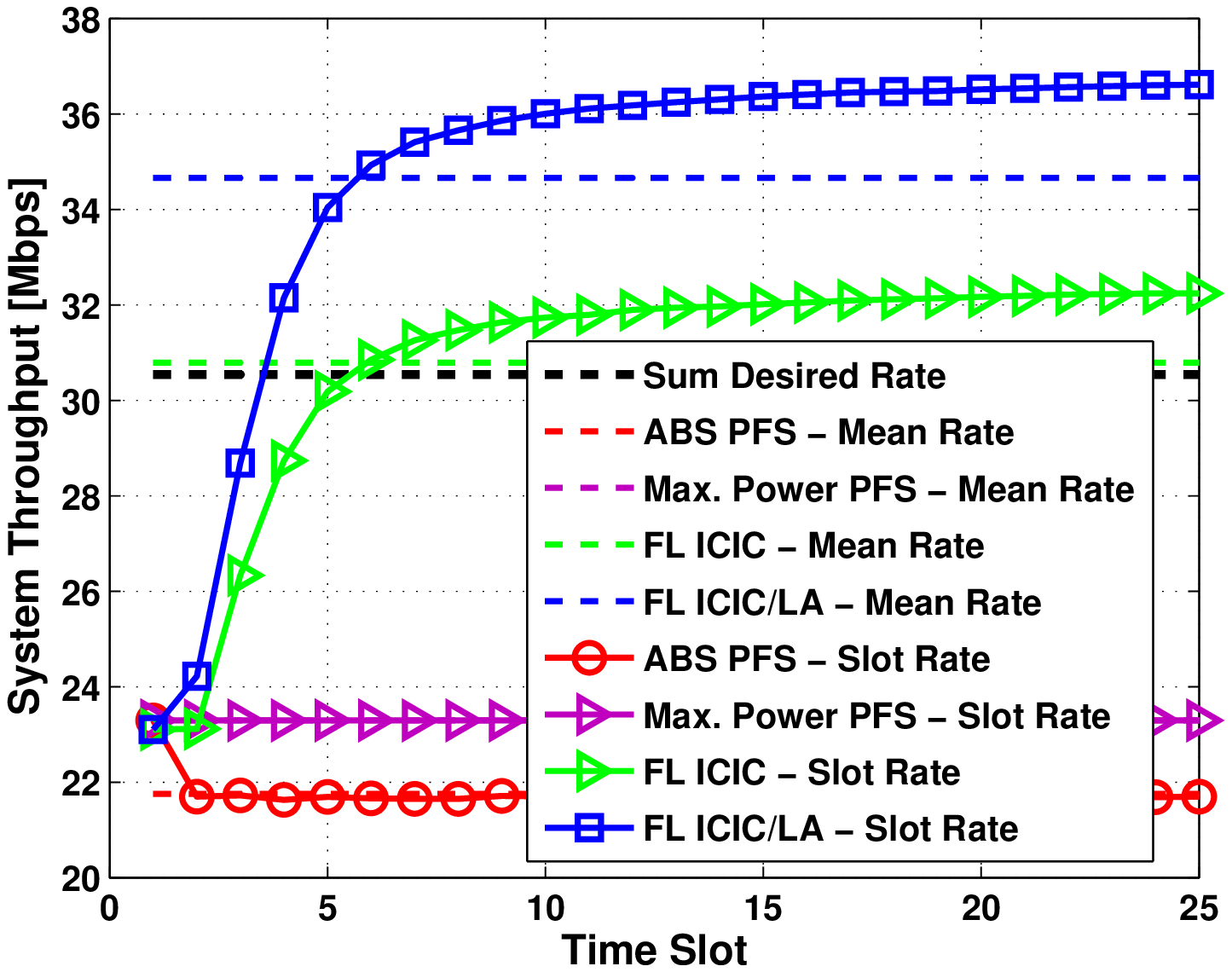}}
\subfigure[Energy Efficiency]{\includegraphics[width=.49\columnwidth]{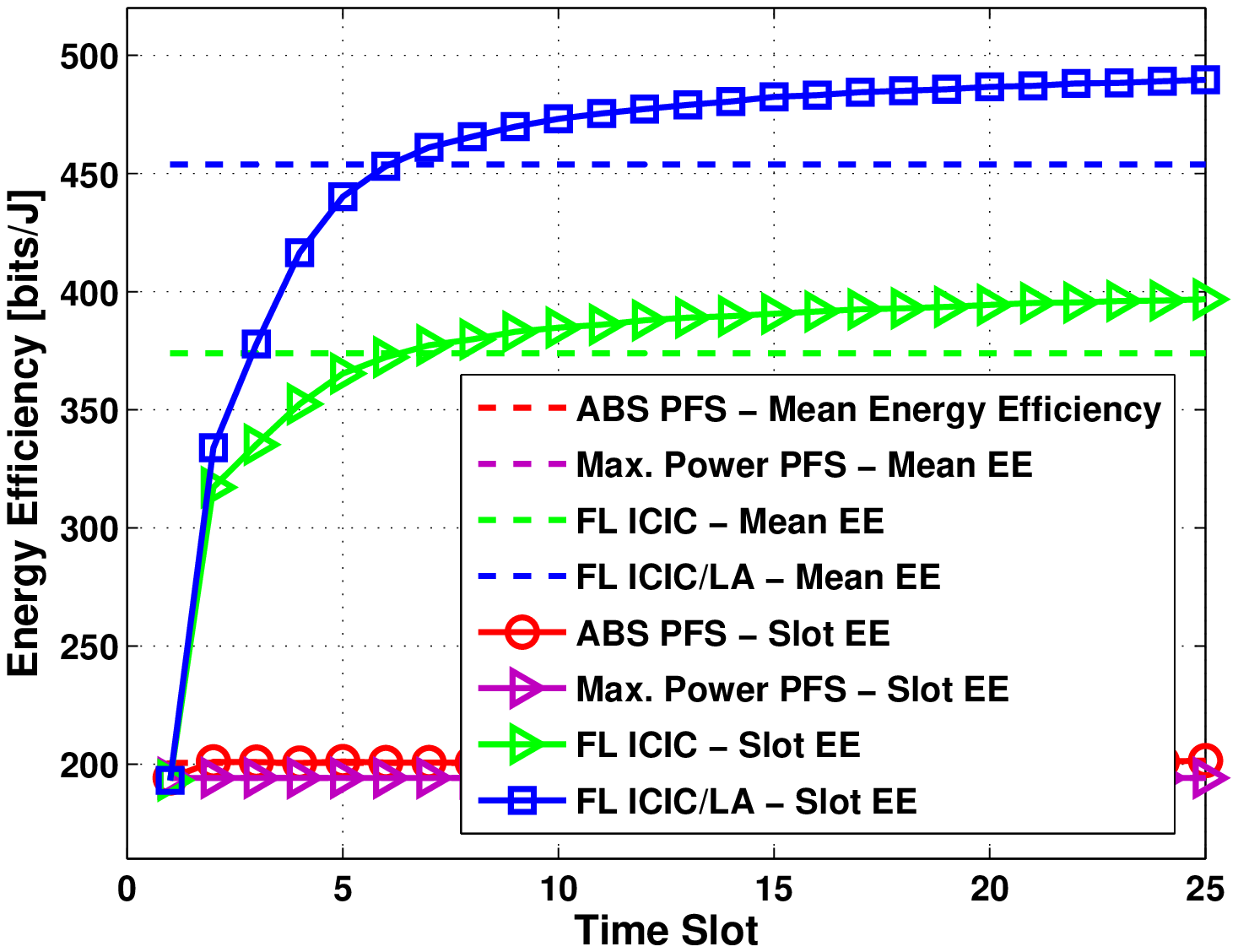}}
\caption{System downlink efficiency performance results of fuzzy logic \ac{icic}, random \acp{abs} transmission, and maximum power transmission.}
\label{fig:results_mp_eff}\vspace*{-20pt}
\end{figure}

It is clear from Fig.~\ref{fig:results_mp_eff} that fuzzy logic \ac{icic} provides substantially improved system performance over both benchmark techniques. Especially in terms of system throughput, where the fuzzy logic schemes are the only techniques which achieve the overall desired rate (\ie sum of individual desired rates). In fact, fuzzy logic substantially overachieves the sum desired rate, indicating almost maximum coverage and all but negligible outage. The additional rate results from the discrete allocation of bandwidth (\ie \acp{rb}), and hence the achieved user rate is generally slightly greater than what was desired. With \ac{la} this becomes even more apparent, as with higher spectral efficiency the throughput ``overshoot'' becomes even greater. 

The \ac{abs} performance is constant over all time slots (except the first), as the probability of \ac{abs} transmission(s) is identical in each slot. 
Hence, in each time slot $10\%$, on average, of the users transmit an \ac{abs}, providing some interference mitigation for the remaining users. This abstinence of data transmission explains the throughput losses by the \ac{abs} system relative to full power transmission, as clearly the interference mitigation provided is less significant than the throughput sacrificed.

Fig.~\ref{fig:results_mp_eff} also displays the energy efficiency of the simulated scenario, yielding again very dominant results of the fuzzy logic systems. This is mainly due to the fact that fuzzy logic has the possibility of transmitting at half power, which is usually the case after multiple time slots and the achievement of a relatively orthogonal \ac{rb} allocation. Furthermore, the high energy efficiency is achieved quite rapidly. 
The added energy efficiency due to \ac{la} is a direct result of the augmented throughputs (see~\eqref{eq:peff}). It is shown that \ac{abs} transmission is slightly more energy efficient than maximum power transmission, which is logical since on average 10$\%$ less power is used, but the loss in throughput is ${<}10\%$, thus enhancing the energy efficiency.
\begin{figure}[hbt]
\centering
\subfigure[Availability]{\includegraphics[width=.49\columnwidth]{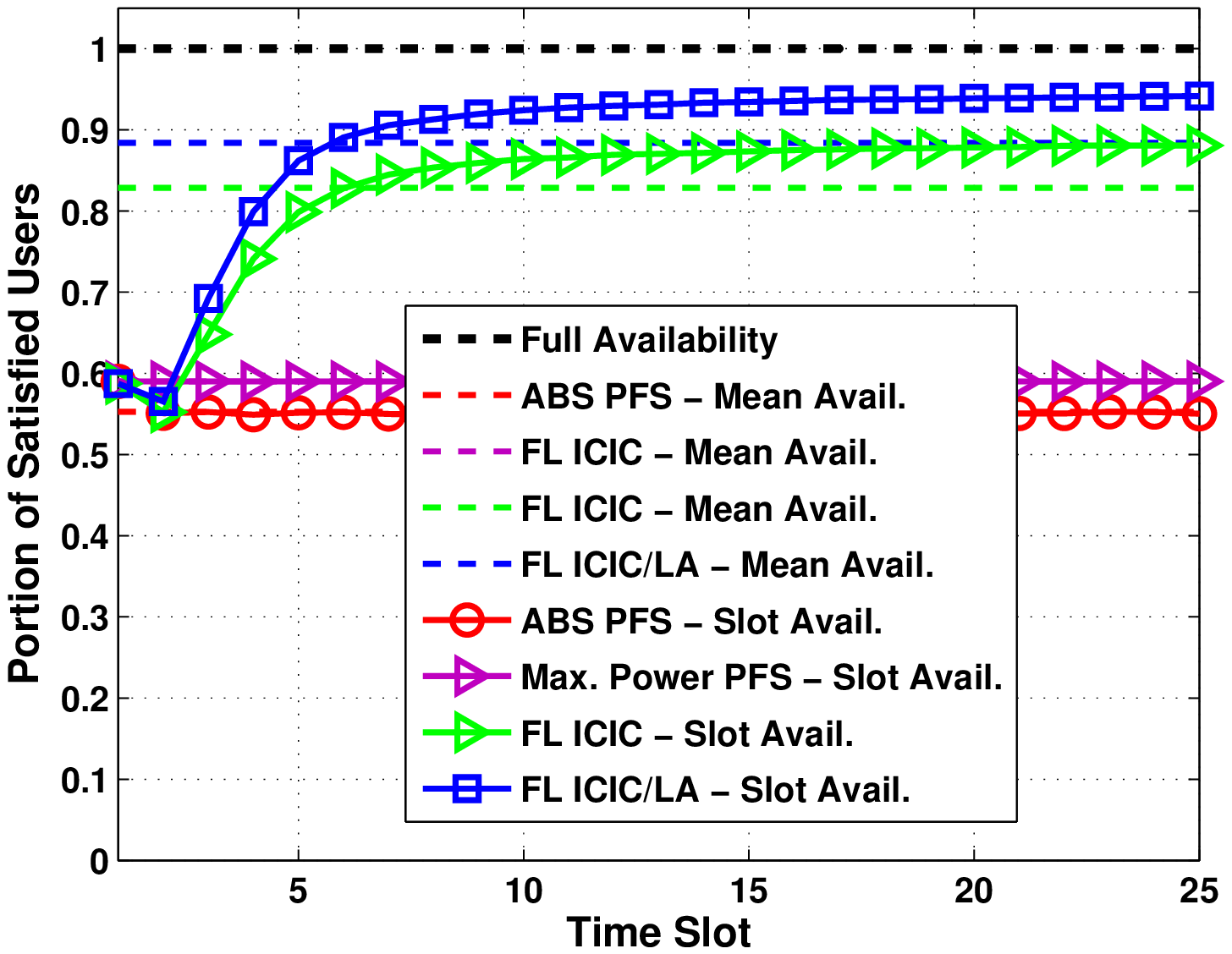}}
\subfigure[Fairness]{\includegraphics[width=.49\columnwidth]{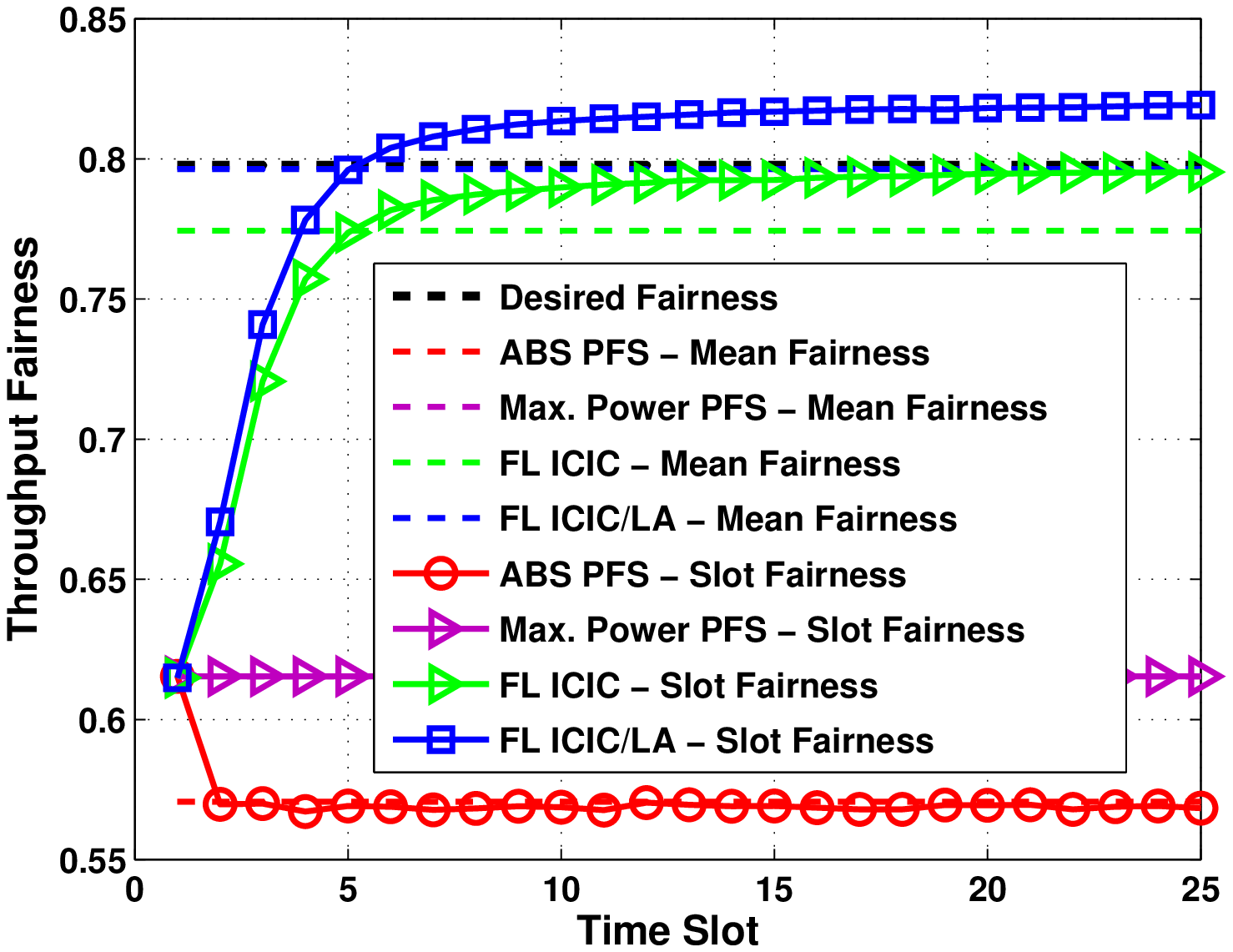}}
\caption{System downlink coverage results of fuzzy logic \ac{icic}, random \acp{abs} transmission, and maximum power transmission.}
\label{fig:results_mp_avail}\vspace*{-10pt}
\end{figure}

Lastly, the availability and throughput fairness in the system are investigated. As expected, fuzzy logic \ac{icic}/\ac{la} provides by far the best \ac{ms} availability, as can be seen from Fig.~\ref{fig:results_mp_avail}, achieving ${\sim}94\%$ availability. This is expected as both the system throughputs are augmented, a direct result of the greater portion of satisfied \acp{ms}. Furthermore, it is clear that the fairness is greatly improved as well, especially when utilising \ac{la}. This is mainly due to the fact that users are (through \ac{la}) more adept to their transmission environments, and hence better achieve their desired rates\footnote{In fact, due to the reduced throughput granularity at higher \acp{mcs}, more \acp{ms} achieve the same throughput, and hence fuzzy logic \ac{icic}/\ac{la} achieves a greater fairness than if all \acp{ms} would exactly achieve their targets.}. On another note, the max. power availability and fairness is boosted with regards to the \ac{abs} system, as all \acp{ms} can transmit without restrictions or abstinence, and hence even unsatisfied (in terms of rate) users achieve decent throughputs. A summary of the quantitative results is shown in Table~\ref{tab:results}.
\begin{table}[htb]
\centering\vspace*{-5pt}
\caption{Performance Results}
\small\vspace*{-5pt}
\begin{tabular}{|c|c|cccc|}
\hline
\textbf{\%-gain} & \textbf{vs.} & \textbf{Throughput} & \textbf{Energy Eff.} & \textbf{Availability} & \textbf{Fairness} \\
\thickhline
FL. ICIC/LA & Max. Power & 57 & 151 & 59 & 33 \\
FL. ICIC & Max. Power & 38 & 103 & 48 & 29 \\
\hline
FL. ICIC/LA & ABS & 68 & 143 & 70 & 44 \\
FL. ICIC & ABS & 48 & 97 & 59 & 40 \\
\hline
FL. ICIC/LA & FL. ICIC & 14 & 24 & 7 & 3 \\
\hline
\end{tabular}
\label{tab:results}
\end{table}\vspace*{-25pt}


\section{Conclusions}
\label{sec:concl}

In this paper, a distributed and autonomous \ac{icic} technique for femto-femto interference management and resource allocation is presented. At each \ac{fbs}, locally available information is utilised to evaluate the allocatability of the available \acp{rb} in a particular cell, taking into account the interference neighbourhood, user rates, and the own-cell signal and fading environment. Fuzzy logic generates broad evaluations of these inputs, combines them based on a defined set of \ac{rb} allocation rules, and submits to the \ac{bs} the most suitable resources and transmit powers for successful and efficient communication. After several time slots and more accurate average signal statistics, the locally optimised resource allocations form a near-optimal global solution.

By comparing fuzzy logic \ac{icic} to a system-wide optimisation problem, it was shown that fuzzy logic provides close-to-optimal system performance with drastically reduced complexity. Furthermore, a comparison to a greedy heuristic of similar complexity shows faster convergence to cell-individual optimum resource allocation. Hence, fuzzy logic provides a low-complexity near-system-optimal solution of \ac{icic} in femto-cell networks. This is confirmed in the simulation results, where fuzzy logic \ac{icic} satisfies the system throughput requirements and significantly outperforms the given benchmarks. The addition of \ac{la} gives a further performance boost, achieving almost full availability along with enhanced throughput, energy efficiency, and fairness. 

The main focus of the further development of fuzzy logic \ac{icic} is the extension to \acp{hetnet}, as highlighted in Section~\ref{sec:intro}. This will see macro-, pico- and femto-cells available in the same scenario, thus the \acp{ms} will not only need to perform resource and power allocation, but also determine which \ac{ap} they desire to connect with. The autonomous and distributed nature of fuzzy logic \ac{icic} should allow these networks to self-configure, and self-optimise, eliminating excessive signalling normally required in such networks. Furthermore, we seek to heuristically optimise the fuzzy logic system (\ie more specifically, the rules) by analysing the input-output characteristics, and tuning the system to make better decisions on each \ac{rb}.

\bibliographystyle{IEEEtran}
\bibliography{../../../reference/cwc,../../../reference/general} 

\end{document}